\documentclass[aps,reprint,
superscriptaddress,floatfix,longbibliography,footinbib]{revtex4-1}
\usepackage{mathtools}
\usepackage{amsfonts}
\usepackage{amsmath}
\usepackage{amssymb}
\usepackage{physics}
\usepackage{float}
\usepackage{bbold}
\usepackage[version=4]{mhchem}
\usepackage{hyperref}
\usepackage{xcolor}
\usepackage{lipsum}
\usepackage{xspace} % only used in \supp, \meth, \xFM, \yFM commands
\usepackage[rightcaption]{sidecap}
\usepackage{graphicx}
\usepackage{tabularray}
\usepackage[dvipsnames]{xcolor}
\usepackage{makecell}

\UseTblrLibrary{booktabs}

\hypersetup{
   colorlinks=true,
   linkcolor=blue,
   filecolor=blue, 
   citecolor=blue,
   urlcolor=blue,
}
\usepackage[doipre={DOI:~}]{uri}

\newcommand{\ex}[1]{{\langle #1 \rangle}}

\newcommand{\mbf}{\mathbf}
\begin{document}

\title{High-capacity associative memory in a quantum-optical spin glass}

\author{Brendan P.~Marsh}
\affiliation{Department of Applied Physics, Stanford University, Stanford, CA 94305, USA}
\affiliation{E.~L.~Ginzton Laboratory, Stanford University, Stanford, CA 94305, USA}
\author{David Atri Schuller}
\affiliation{Department of Applied Physics, Stanford University, Stanford, CA 94305, USA}
\affiliation{E.~L.~Ginzton Laboratory, Stanford University, Stanford, CA 94305, USA}
\author{Yunpeng Ji}
\affiliation{Department of Applied Physics, Stanford University, Stanford, CA 94305, USA}
\affiliation{E.~L.~Ginzton Laboratory, Stanford University, Stanford, CA 94305, USA}
\affiliation{Department of Physics, Stanford University, Stanford, CA 94305, USA}
\author{Henry S.~Hunt}
\affiliation{E.~L.~Ginzton Laboratory, Stanford University, Stanford, CA 94305, USA}
\affiliation{Department of Physics, Stanford University, Stanford, CA 94305, USA}
\author{\\Surya Ganguli}
\affiliation{Department of Applied Physics, Stanford University, Stanford, CA 94305, USA}
\author{Sarang Gopalakrishnan}
\affiliation{Department of Electrical and Computer Engineering, Princeton University, Princeton, NJ 08544, USA}
\author{Jonathan Keeling} 
\affiliation{SUPA, School of Physics and Astronomy, University of St. Andrews, St. Andrews KY16 9SS, United Kingdom}
\author{Benjamin L.~Lev}
\affiliation{Department of Applied Physics, Stanford University, Stanford, CA 94305, USA}
\affiliation{E.~L.~Ginzton Laboratory, Stanford University, Stanford, CA 94305, USA}
\affiliation{Department of Physics, Stanford University, Stanford, CA 94305, USA}

\date{\today}

\begin{abstract}  

The Hopfield model describes a neural network that stores memories using all-to-all-coupled spins. Memory patterns are recalled under equilibrium dynamics. Storing too many patterns breaks the associative recall process because frustration causes an exponential number of spurious patterns to arise as the network becomes a spin glass. Despite this, memory recall in a spin glass can be restored, and even enhanced, under quantum-optical nonequilibrium dynamics because spurious patterns can now serve as reliable memories. We experimentally observe associative memory with high storage capacity in a driven-dissipative spin glass made of atoms and photons. The capacity surpasses the Hopfield limit by up to seven-fold in a sixteen-spin network. Atomic motion boosts capacity by dynamically modifying connectivity akin to short-term synaptic plasticity in neural networks, realizing a precursor to learning in a quantum-optical system.

\end{abstract}

\maketitle

Content-addressable associative memories can convert corrupted or incomplete data into clean, stored memory patterns through the process of pattern completion. For example, one may want to remember the face of a friend based on a blurry photo. Successful pattern completion outputs the unblurred image. (Too much blurring may result in the recall of a different friend.)  Ordered states of matter can store such memories: For example, an Ising ferromagnet encodes one bit of information, namely whether the spins are mostly up or down. This encoding is highly redundant, and therefore robust against errors, but stores only one pattern as a memory, the spins-all-aligned state. Simultaneously maximizing both memory storage capacity and robustness to recall error has wide-ranging implications from the fields of artificial intelligence (AI), where dense associative memories have been closely linked to the transformer architecture that now dominates AI, particularly large language models~\cite{Ramsauer2020hni}, to neuroscience, where such associative systems are long thought to mediate human episodic memories~\cite{Marr1971sma}.

The Hopfield model, a subject of the 2024 Nobel Prize in Physics, is an example of a physically motivated associative memory that can store multiple patterns~\cite{Hopfield1982nna,Hopfield1986cwn,Bialek2024mba}. It employs a recurrent neural network based on Ising spins that can i)~learn and store new memories through a biologically plausible mechanism (Hebbian learning~\cite{Hebb1949too}), and ii)~effect pattern completion via equilibrium dynamics such as Metropolis-Hastings (MH) dynamics~\cite{Metropolis1953eos}. These dynamics consist of a stochastic update that corresponds to equilibrium energy exchange with a heat bath. The network consists of $n$ binary variables (called ``spins'' or ``neurons'') $s_i = \pm 1$. The energy of a spin configuration is $E = -\sum\nolimits_{ij} J_{ij} s_i s_j$, where the Hebbian coupling weights (or ``synapses") are $J_{ij} = \sum_{p = 1}^P \xi_i^p \xi_j^p$ and each $n$-dimensional binary vector $\xi_i^p$ is a stored pattern, i.e., one of the intended stored memories. 

\begin{figure*}[t]
    \centering
    \includegraphics[width=0.85\linewidth]{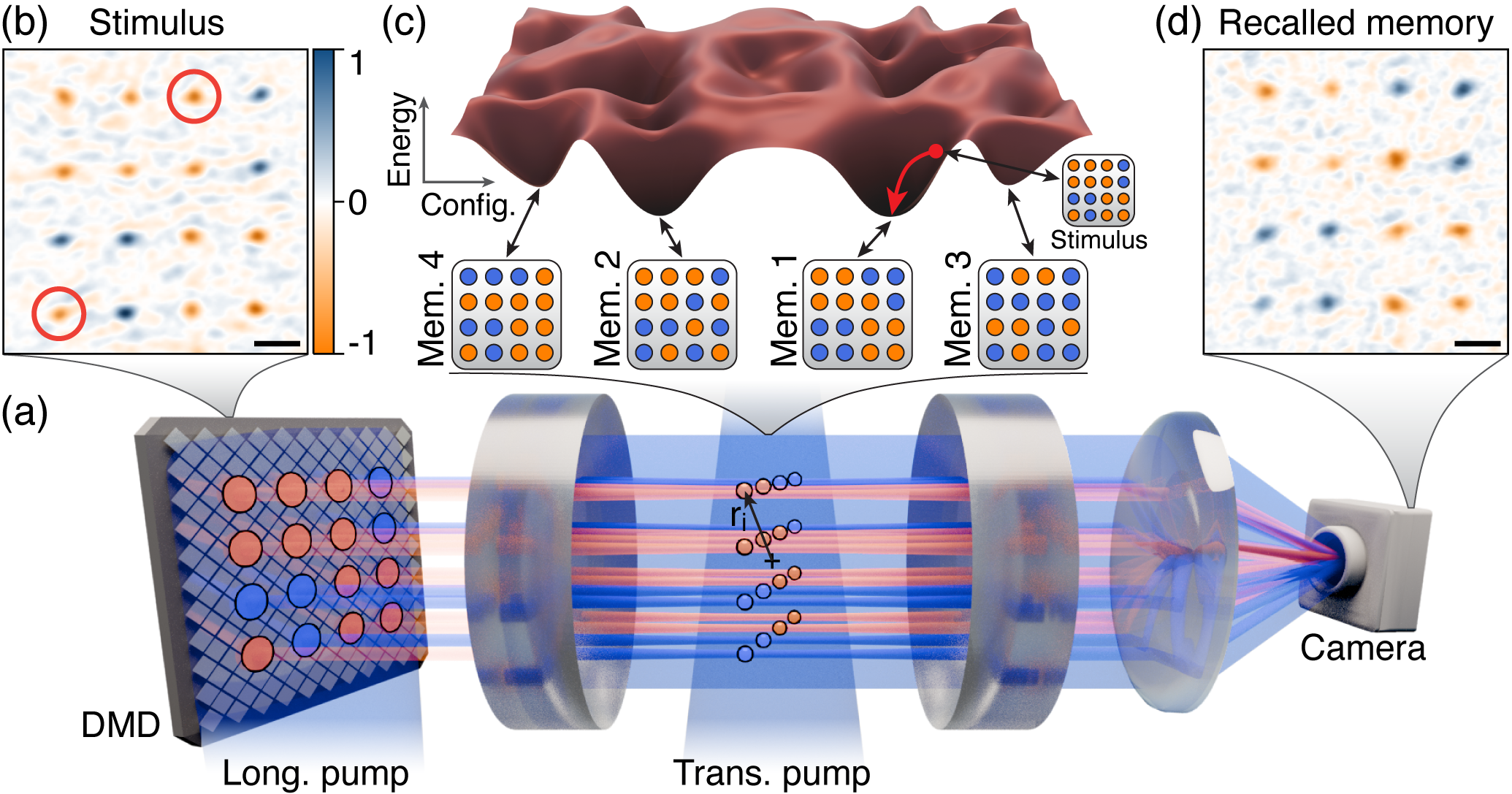}
    \caption{Sketch of the apparatus and memory recall process for an $n=16$ network.   a) Transverse and longitudinal fields (light blue) pump the cavity.  A digital micromirror device (DMD) scatters the longitudinal pump into 16 beams of phase 0 ($\pi$) colored  blue (orange). Each imposes a longitudinal field $f_i$ onto a single atomic ensemble at position $\mbf{r}_i$ within the cavity midplane, thereby inputting into the network a stimulus field.  (Tweezer traps not shown.) The intracavity field contains both local (blue and orange) and nonlocal (light blue) components. Atomic ensembles are represented as spheres. They are colored blue (orange) to denote their effective spin up (down) states, which scatter into local fields with 0 ($\pi$) phase. A camera holographically images the emitted cavity field~\cite{Guo2019spa,Marsh2025amc}.  b) Example stimulus field pattern. Spin-flip errors compared to memory~1 are circled in red. Image taken by recording the transmission of the longitudinal pump beams through an empty cavity. Images are processed to show only local fields and normalized to the their maximum value. Remaining background is residual noise; see~\cite{Marsh2025amc} for image analysis. Black bars are of length $w_0 = 34.8$~$\mu$m.  c) Memory recall dynamics in a low-dimensional representation of a spin-glassy energy landscape versus spin configuration. Red arrow indicates gradient descent from the corrupted stimulus pattern (red dot) in panel (b) to the memory pattern 1 at the bottom of its basin of attraction. Three additional representative memory patterns are indicated at their respective basin minima.
    d) Cavity output image showing the local fields from the atomic ensembles after successful recall of memory 1, starting from the stimulus field in panel (b).
    }
    \label{Fig1}
\end{figure*}

By analogy with the ferromagnet, one might expect that memories are retrievable in the Hopfield model so long as the system is in a (generalized) ferromagnetic phase, where the $P$ patterns $\xi_i^p$ generalize the all up or down ``pattern'' in the conventional ferromagnet. Indeed, Amit \textit{et al.}~showed that such ferromagnetic retrieval states exist for $P$ less than the thermodynamic bound  $P_0 \approx 0.14 n$~\cite{Amit1985sin}. In this regime, the patterns $\xi_i^p$ form local minima of $E$, and pattern completion occurs via stochastic energy descent to those minima. However, when $P \geq P_0$, the retrieval states become thermodynamically unstable and an equilibrium spin glass forms~\cite{Stein2013sga,Charbonneau2023sgt}. This hampers recall through the formation of a rugged energy landscape with exponentially many local minima residing in deep, nested valleys separated by high energy barriers. Most minima do not coincide with any intended pattern $\xi_i^p$ and are thus ``spurious." Moreover, the spurious patterns cannot serve as reliable memories under MH dynamics: An error can lead to other patterns in nearby energy valleys. For this reason, spin glasses have long been considered incompatible with associative memory.

However, a recent theoretical work suggested that switching to a particular form of nonequilibrium dynamics would open access to the exponentially larger storage capacity of glassy systems~\cite{Marsh2021eam}. Driven-dissipative dynamics in a system of atoms and photons were theoretically found to induce energy-lowering spin flips at a rate proportional to the energy lost upon flipping.  This realizes deterministic, ``steepest descent" dynamics that reduces the multiplicity of possible relaxation pathways, thereby enlarging the basin size of any given spurious pattern. (By contrast, under MH dynamics, any spin-flip that lowers the energy is equally likely to happen.)  Decreased entropy generation due to steepest descent dynamics enhances recall fidelity, thereby turning what would have been spurious patterns under equilibrium recall dynamics into high-fidelity memories under steepest descent dynamics. That is, the spurious pattern, residing at the minimum of a basin of attraction, can now be considered a natural memory of the network defined by the spin-connectivity weights $J_{ij}$.

We experimentally demonstrate the enhanced memory capacity of a spin glass using an all-to-all Ising spin network created with multimode cavity QED~\cite{Marsh2025amc}.  Driven-dissipative dynamics are natural to cavity QED-based quantum-optical systems, and the theoretical possibility for creating associative memory therein has been discussed~\cite{Gopalakrishnan2011fag,Gopalakrishnan2012emo,Rotondo2018oqg,Marsh2021eam}. Not discussed, however, is a mechanism we report here by which the $J_{ij}$ evolve as dynamical quantities that depend on spin position and contributes to capacity enhancement. While many optics and nonlinear optics experiments have realized associative memory~\cite{Xu2021aso,Shastri2021pfa,Katidis2025rpr,Kalinin2025aoc}, none have observed memory enhancement due to spin glass ordering. ``Dense" associative memory networks can also possess enhanced capacity, though at the expense of engineering greater-than-2-body spin interactions~\cite{Krotov2016dam,Musa2025dam}.   We also note that photonic neural networks at the few-quanta level are under development~\cite{Ma2025qso}.  

We perform associative memory recall in artificial neural networks of size up to $n = 20$ and extensively characterize networks with $n\leq 16$.  Bose-condensed gases of atoms play the role of the spins (neurons) while photons resonating in an optical cavity mediate the synaptic connections ($J_{ij}$ weights).  The memory capacity can exceed the thermodynamic bound of the Hopfield model $P_0$ by up to seven-fold at $n=16$,
using a 50\% recall probability threshold; see  supplementary materials~\cite{Supp} for a discussion of this threshold choice.  Moreover, this is larger than the capacity limit equal to $n$ attained by replacing Hebbian learning with a pseudoinverse learning rule~\cite{Storkey1999tbo}.

The high memory capacity is attributed to two experimental features that are absent in the Hopfield model.  First, our results are consistent with theoretical expectations~\cite{Marsh2021eam} (albeit derived in a different parameter regime) that a cavity-cooling mechanism~\cite{Vuletic2000lco} intrinsic to the experiment contributes to deterministic spin relaxation, rather than a solely stochastic energy descent. Second, the $J_{ij}$ depend on the spin positions $\mbf{r_i}$ that are not frozen in space, but can respond elastically to a ``synaptic" stimulus. This renders $\mbf{J}$ a dynamical quantity through a spin-motion coupling $J_{ij}=J(\mbf{r}_i,\mbf{r}_j)$ and stabilizes memory patterns. In response to optical forces generated by the spin-dependent cavity field, each atomic gas shifts its position to perturb its connectivity $J(\mbf{r}_i,\mbf{r}_j)$ with all others. The modified $J_{ij}$ can then flip spins to drive the system deeper into a basin of attraction. The cavity field is modified with each spin flip, inducing more movement, resulting in a self-reinforcing interplay between spin and motional degrees of freedom until the system evolves to a configuration deep within a basin of attraction.  In consequence, the energy landscape elastically deforms under the driven-dissipative cavity dynamics to assist the spin evolution toward the memory. Simulations presented in Ref.~\cite{Supp} replicate this elastic response and indicate that this enhances memory capacity.  Thus, we realize a form of ``polaronic spin glass," one whose connectivity dynamically changes to provide self-reinforced memory recall. This is akin to the creation of polarons in crystal lattices where the position of a charge leads to a deformation of the lattice, which in turn acts to trap the charge~\cite{Franchini2021pim}.

Our elastic enhancement of memory is analogous to ``short-term synaptic plasticity" in neurobiology~\cite{Zucker2002ssp}, where transient facilitation of synapses can enhance memory capacity~\cite{Mejias2009mmc}. This enhancement arises through transiently altered synaptic connectivity that deepens and reinforces the energy basin in which neural activity currently resides.  While the training of physical neural networks has been considered~\cite{Momeni2025top}, the utilization of synaptic self-reinforcement through natural dynamics in magnetic and photonic systems remains in early-stage development~\cite{Feldmann2019asn,Niu2024asm}.

\begin{figure}[t!]
    \centering
    \includegraphics[width=\linewidth]{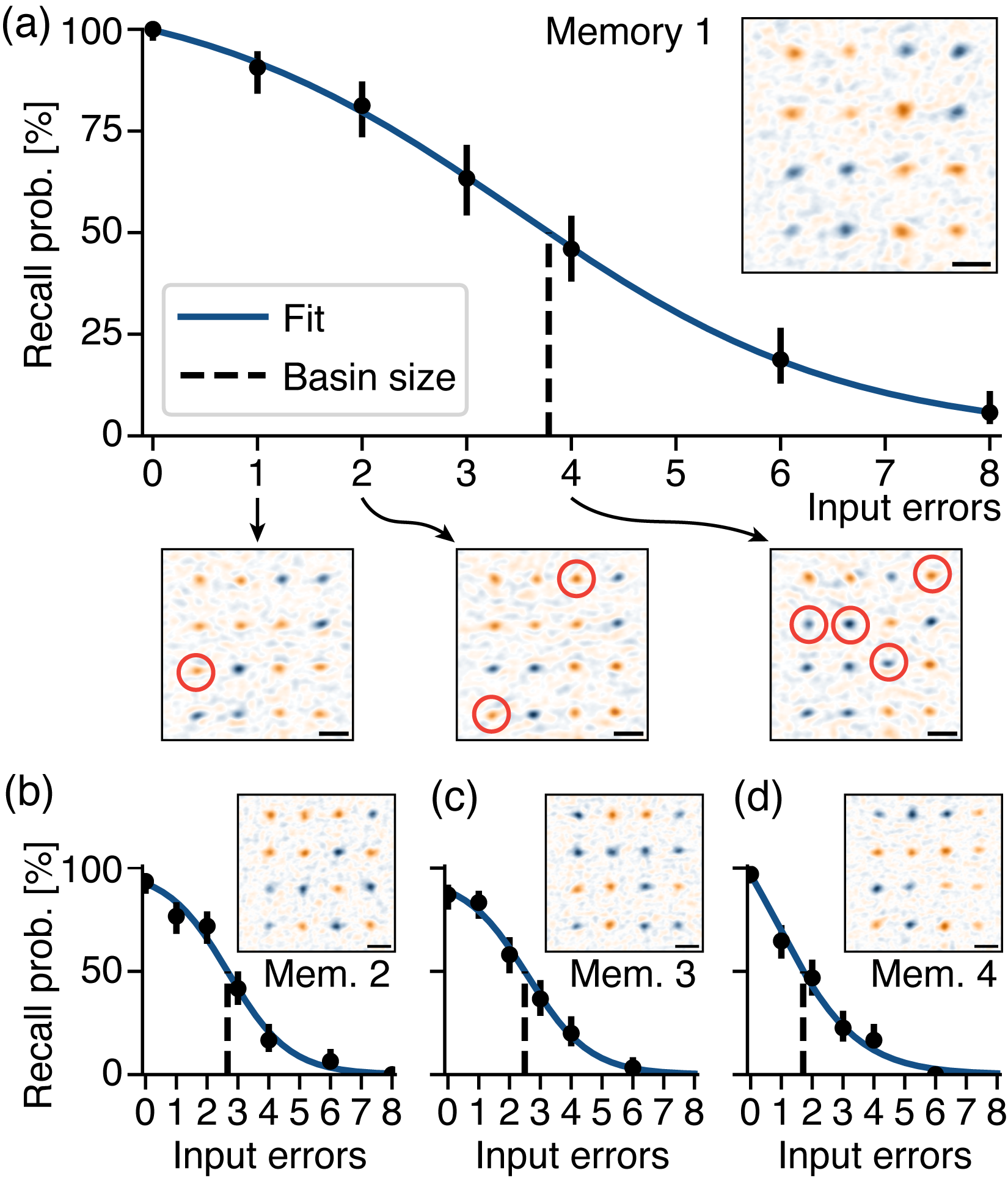}
    \caption{Memory recall fidelity.  a) Recall probability versus number of spin-flip errors in memory pattern 1 of connectivity realization $\mathbf{J}_1$, shown as inset (same image as in Fig.~\ref{Fig1}d).   Below are examples of three defective stimulus images including one, two, and four spin-flip errors, circled in red. The dashed line shows the basin size, which is chosen where the fitted curve (blue line) intersects the 50\% probability threshold; a hyperbolic tangent is empirically found to be a suitable fitting function~\cite{Supp}. b-d) Recall curves for three additional memories of $\mathbf{J}_1$.  Insets are emission images showing the memory patterns.  Black scale bars are of length $w_0$.}
    \label{Fig2}
\end{figure}

The experimental system has been discussed in Refs.~\cite{Kroeze2025dor,Marsh2025amc}; see~\cite{Supp} for parameters and procedures specific to this work. Figure~\ref{Fig1}a sketches the apparatus, which has been augmented to enable the application of bias fields onto each spin.  Briefly, each spin is represented by the motional excitation of a small rubidium Bose-Einstein condensate (BEC) trapped by an optical tweezer. These motional states are one of two possible checkerboard density waves representing collective spin up or down states. This yields an effective Ising spin degree of freedom $\hat{S}_i^x$~\cite{Marsh2025amc}. The density wave states are created by the interference of a transverse pump laser and the field of the multimode cavity within which the BECs are confined~\cite{Baumann2010dqp,Kollar2015aac,Mivehvar2021cqw}. 

BECs in spin up (down) density-wave states scatter the pump light into the cavity with phase 0 ($\pi$) with respect to this transverse pump.  Recording the phase and amplitude of the cavity emission from each of these BECs using holographic imaging allows us to measure all $\ex{\hat{S}_i^x}$ of a microscopic spin configuration~\cite{Guo2019spa,Kroeze2023hcu,Marsh2025amc}; e.g., see Fig.~\ref{Fig1}d. Longitudinal pump beams provide local fields $f_i$ that bias these effective spins toward a stimulus pattern $\ex{\hat{S}_i^x}\propto f_i$; see Fig.~\ref{Fig1}b. (We note that rarely, a BEC splits into two opposite density waves of unequal weight. This does not affect our ability to determine the spin configuration; see Refs.~\cite{Supp,Marsh2025amc} for explanation.) 

The multimode cavity light---comprised of thousands of near-degenerate modes~\cite{Kollar2015aac,Kroeze2023hcu}---also induces all-to-all, sign-changing interactions among the effective spins~\cite{Gopalakrishnan2011fag,Guo2019spa,Guo2019eab}. When the BECs are spaced far enough from both the cavity center and each other, the interaction yields a frustrated $J_{ij}$ connectivity matrix sufficient to create a spin glass~\cite{Marsh2021eam}.  See Refs.~\cite{Supp,Marsh2025amc} for the form of this interaction and Refs.~\cite{Kroeze2025dor,Marsh2025amc} for measurements of the ``overlap'' order parameters~\cite{Stein2013sga} that characterize this intrinsically nonequilibrium spin glass~\cite{Kroeze2025dor}---i.e., one that is coherently driven while under the influence of dissipation through cavity emission.

We previously showed that the multimode cavity-mediated interactions yield a spin-glass Hamiltonian of Ising form when the cavity is tuned to the so-called ``4/7" configuration of a Fabry-P\'{e}rot resonator~\cite{Marsh2025amc}. The 4/7 cavity geometry is realized when the ratio of mirror spacing $L$ to radius of curvature $R$ is $\sim$1.22; here, $R = 1$~cm. The spin-glass Hamiltonian arises in the dispersive pumping limit, wherein the pumps are red-detuned $\Delta_C\approx -2\pi{\cdot}20$~MHz from the near-degenerate 4/7 cavity modes.  The multimode Dicke model description of the system then simplifies to a frustrated, transverse-field Ising model with longitudinal fields~\cite{Supp}:
\begin{equation}\label{hammain}
    \hat{H} = \omega_z \sum_{i=1}^n \hat{S}_i^z - g\sum_{ij=1}^n J(\mbf{r}_i,\mbf{r}_j)\hat{S}_i^x \hat{S}_j^x - \sum_{i=1}^n f_i \hat{S}_i^x.
\end{equation}
Each spin is represented by a collective spin operator $\hat{S}_i^{x/y/z}$ of size $S = M/2$, where $M\approx 4{\cdot}10^4$ is the number of atoms per ensemble, on average.  The form of $J(\mbf{r}_i,\mbf{r}_j)$ is in~\cite{Supp}.  The experimental dynamics may be approximated by unitary evolution through Eq.~\eqref{hammain} combined with dissipation using a Lindbladian treatment~\cite{Marsh2021eam,Supp}. The term proportional to the atomic recoil frequency $\omega_z\approx 2\pi{\cdot} 7.5$~kHz plays the role of a transverse field.  The strength of the interaction term is $g=-g_0^2\Omega^2/(\Delta_A^2\Delta_C)$, where the Rabi rate squared $\Omega^2$ is proportional to the transverse pump power, $g_0 = 2\pi{\cdot} 1.35$~MHz is the Rb atom-cavity coupling strength, the detuning of the pumps from the atomic transition is $\Delta_A\approx -2\pi{\cdot}97.2$~GHz, and the cavity linewidth is $\kappa = 2\pi{\cdot}140$~kHz.  The system is in the single-atom, large cooperativity limit of cavity QED, even without multimode field enhancement~\cite{Kroeze2023hcu,Supp}.

Associative memory recall is performed as follows.  The transverse pump is exponentially increased in power till the spin interactions are as large as $2\pi{\cdot}$2~kHz at 8~ms. Simultaneously, a memory stimulus is input into the network by ramping up the longitudinal fields over 3~ms. This induces $|f_i|$ whose strengths are approximately $\omega_z$. Figure~\ref{Fig1}b shows an example stimulus pattern for the connectivity matrix we call $\mathbf{J}_1$; see~\cite{Supp} for all $J_{ij}$ elements. The transverse and longitudinal pumps drive spin evolution from the stimulus configuration down toward lower-energy configurations within a basin of attraction, as depicted in Fig.~\ref{Fig1}c.  The biasing fields are held constant for another $\sim$3~ms to allow spin organization to continue at fixed $|f_i|$. We then ramp down $f_i$ before turning it off 1-ms before imaging to allow unbiased spin evolution toward a memory configuration; these are attractors of the dynamics. In this work, we do not explore long-term aging dynamics in the glassy landscape. An example of successful pattern recall is shown in Fig.~\ref{Fig1}d for what we designate as memory 1 of $\textbf{J}_1$. Recall fidelity is qualitatively insensitive to small changes in this procedure~\cite{Supp}. To operate the neural network again, fresh BECs are created and positioned to realize the same $\textbf{J}$, within experimental uncertainty~\cite{Marsh2025amc,Supp}.

Different coupling matrices $\mathbf{J}$ realize different neural networks, each with their own memories.  We study five disorder realizations of $\mathbf{J}$ for each system size $n=4, 8, 12$ and 16. To obtain different $\mathbf{J}$'s at fixed $n$, we simply move the ensembles within the cavity midplane. This causes the $J_{ij}$ to change sign in a sufficiently random manner that the resulting glassy energy landscapes store distinct sets of memory patterns~\cite{Marsh2021eam,Kroeze2025dor,Marsh2025amc}. In practice, it suffices to trap the ensembles in a rectilinear array of spacing roughly $50$~$\mu$m from one another, and few-micron-scale adjustments to the row and column locations yield significantly different $\mathbf{J}$'s~\cite{Supp}. We report memories recalled from a single $\mathbf{J}$ at $n=20$ in~\cite{Supp}.  Note that the rapid proliferation of memories renders capacity measurements prohibitively time consuming for $n>16$, at present.

Memory capacities are determined by cataloging all memories with finite basins of attraction that are naturally stored by each $\mathbf{J}$. That is, the $\mathbf{J}$ are not trained to realize specific patterns in this work. Rather, we sample the patterns found by performing up to 400 experimental recall cycles, each with the same $\mathbf{J}$ but a randomly selected stimulus pattern $\{f_i\}$. A hierarchical clustering algorithm~\cite{Bar-Joseph2001fol} organizes the observed patterns into groups of similar patterns to allow for small fluctuations in $\ex{\hat{S}_i^x}$ between experimental shots~\cite{Supp}. Each group serves as a ``candidate memory." We then use the most commonly found pattern in each group as the reference pattern to measure the basin size~\cite{Supp}.

We define the basin size of a candidate memory as the average number of randomized spin flip errors that may be tolerated in the stimulus while exceeding a 50\% recall probability threshold. We find an average of 51 candidates for the five $\mathbf{J}$'s studied at $n=16$. However, most candidates exhibit basin sizes smaller than one spin flip and are therefore not useful as associative memories. This is expected:  Reference~\cite{Marsh2021eam} predicted a typical basin size of the spin glass to be ${\sim}0.013n$, which is less than one for $n=16$. Thus, for small $n$, it is reasonable to restrict ``memories" to mean only those memory candidates with basin sizes greater than or equal to one spin flip. The capacity we quote for each $\mbf{J}$ is the number of memories that satisfy this criterion. We estimate that we find $>$95\% of all such memories for $n\leq16$~\cite{Supp}. 

Figure~\ref{Fig2} shows recall fidelity curves for representative memories of $\mathbf{J}_1$.  Memory 1 in panel (a) exhibits a particularly large basin size of nearly four spin flips. Figures~\ref{Fig2}b-d present typical recall curves for three additional memories of $\mathbf{J}_1$, with basin sizes between 1-3 spin flips. We find $\mathbf{J}_1$ has a memory capacity of 14(1) memories~\cite{Supp}. See~\cite{Supp} for a gallery of the ten other memories.

\begin{figure*}[t]
    \centering
    \includegraphics[width=\linewidth]{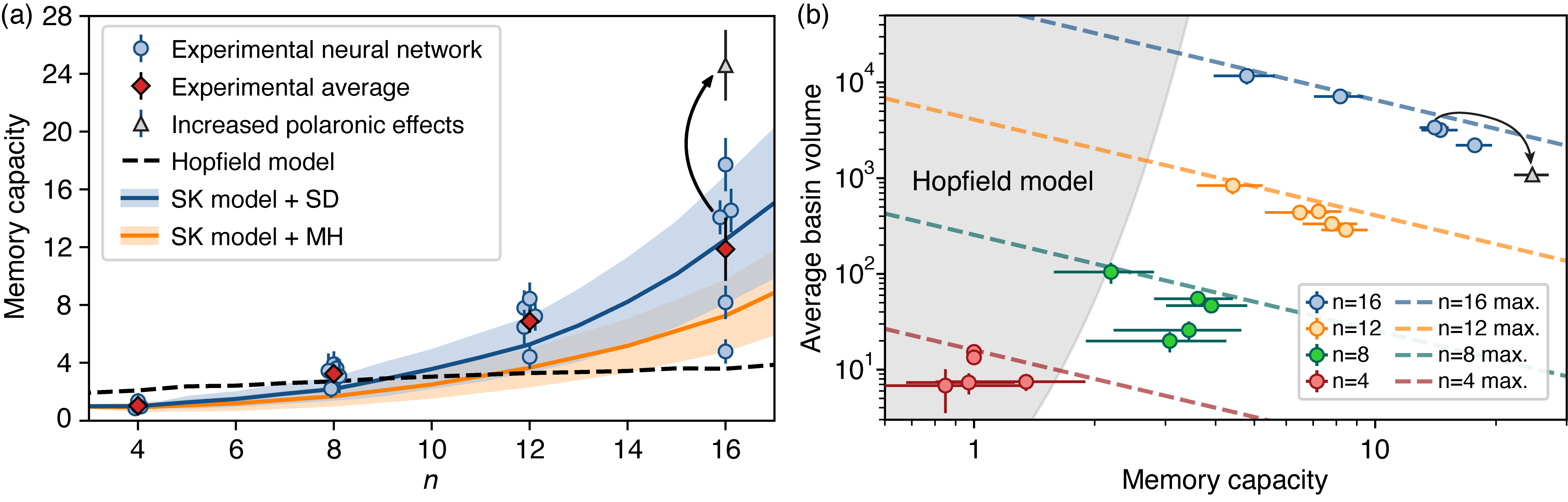}
    \caption{Memory capacity and average basin volume versus $n$.   a) Memory capacity of five different $\textbf{J}$'s for each of four different system sizes $n$. Their average is shown as red diamonds. Error bars are standard error and stem from basin size estimation~\cite{Supp}.  Here and in panel (b), arrows indicate the polaronic enhancement of memory capacity for the $\textbf{J}_1$ network, where capacity is measured at both the nominal trap elasticity condition (blue circles) and at a higher elasticity (gray triangle). The solid curves show the simulated memory capacity in the SK spin glass under steepest descent (SD, blue) or Metropolis-Hastings (MH, orange) dynamics. The dashed black curve is for the Hopfield model.  The SK (Hopfield) simulations include 1,000 (10,000) independent disorder realizations of $\mbf{J}$ per $n$.  The light blue (orange) band shows the standard deviation in SD (MH) capacities due to random variations in the disordered energy landscape versus $\textbf{J}$. This band is ${\sim}1.3$ memories for the Hopfield model~\cite{Supp}. 
    b) Comparison of the measured average basin volumes to the upper bounds (dashed lines) at each $n$. Capacity limits restrict Hopfield networks to lie within the shaded region.  
    }
    \label{Fig3}\vspace{-4mm}
\end{figure*} 

We measure the capacity for all $\mathbf{J}$ and $n$ and plot these in Fig.~\ref{Fig3}a. The average capacity 11.9(6) of the $n=16$ network is much greater than the 3.6 average number of memories storable by the Hopfield model~\cite{Supp}. As expected, however, the average basin size of the memories of $\mathbf{J}_1$ is 2.1, which is lower than that of the Hopfield model, 3.9. This highlights the trade-off noted in Ref.~\cite{Marsh2021eam} between a spin glass's larger capacity and its smaller average basin sizes; ultimately, specific application requirements may prefer one over the other. 

Plotting the number of natural memories in the Sherrington-Kirkpatrick (SK) model~\cite{Sherrington1975smo,Stein2013sga} allows us to compare to a model that is known to possess an exponential number of minima in $n$~\cite{Binder1986sge}. (The SK model describes an all-to-all spin glass similar to what we realize; see~\cite{Marsh2021eam,Kroeze2025dor,Marsh2025amc} for a discussion.)  We find that the mean experimental capacity is consistent with the SK model under steepest descent (SD) dynamics (up to $n=16$). Moreover, their spreads in capacity versus disorder realization $\mathbf{J}$ are consistent.  By contrast, the SK model simulated under MH dynamics does not match. While we cannot yet directly prove that purely steepest descent dynamics are at play, recording the spins' temporal evolution would provide direct evidence. We further note that the match with SD dynamics may provide only a lower bound to the possible enhancement to memory capacity for two reasons: 1) As we discuss below, ensemble position elasticity can double the memory capacity; and 2) simulations in section IV.C~\cite{Supp} show that eliminating the DMD phase noise on the stimulus field can improve capacity by 2-3$\times$,  pointing to a straightforward technical path toward further capacity enhancement.  

We also note that one can quantify the recall performance through the basin volume, defined as the number of spin configurations that flow to a given memory pattern. An ideal associative memory utilizes the full spin-configuration space, such that all stimulus patterns flow to memories possessing high recall fidelity. This corresponds to an upper bound on the average basin volume per memory of $2^n$ divided by the memory capacity. We find that experimentally derived basin volumes, shown in Fig.~\ref{Fig3}b, come close to saturating this bound.

\begin{figure}[t]
    \centering
    \includegraphics[width=\columnwidth]{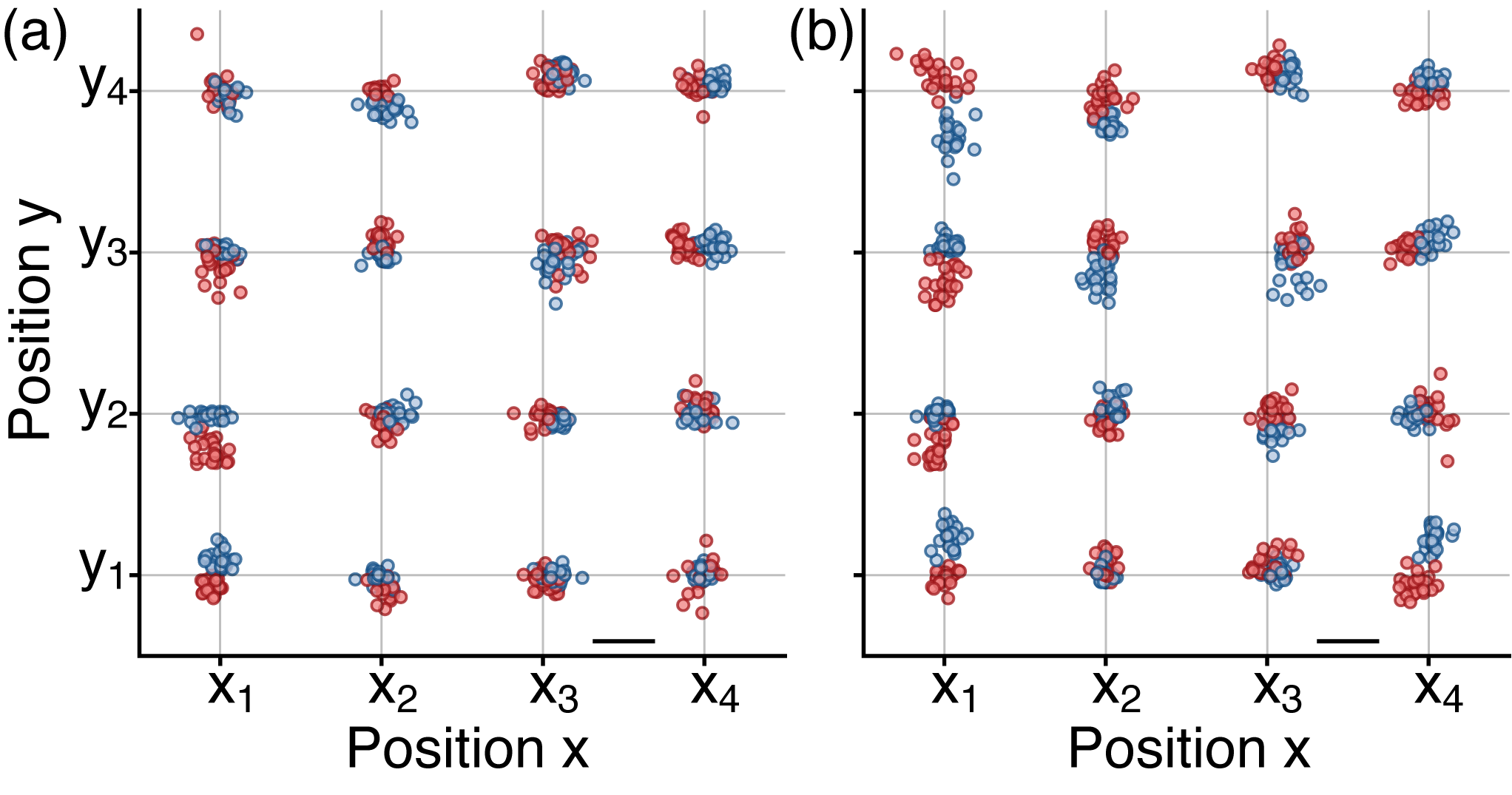}\vspace{-3mm}
    \caption{Polaronic response of atomic ensemble positions. a) Each dot is the center-of-mass (CoM) position of an ensemble within the $4{\times}4$ array of the $n=16$ network. Grid lines at $x_i$ and $y_j$ indicate the tweezer trap locations. The scale bar is 5~$\mu$m.  Positions of the ensembles at the time of imaging are shown in blue (red) after successful recall of memory pattern 1 (2) of $\mathbf{J}_1$.  These are determined by fits to recall images like that shown in Fig.~\ref{Fig1}d. The CoM data from a total of 24 experimental shots are displayed per memory pattern. The average deviation of the CoM data from the trap locations is $1.32(2)$~$\mu$m. b) Same as panel (a), but with trap power reduced four-fold to enhance polaronic elasticity, corresponding to the triangle data point in Fig.~\ref{Fig3}. The average deviation increases to $1.82(3)$~$\mu$m. Note that the spacings between adjacent $x_i$ (and $y_j$) are decreased by $\sim$60\% to visually magnify the spread in positions.}
    \label{Fig4}\vspace{-4mm}
\end{figure}

Experimental observations additionally reveal a new aspect of the dynamics not considered in Ref.~\cite{Marsh2021eam}: We notice that the atomic ensembles shift in position depending on which memory pattern is recalled. This aforementioned polaronic effect occurs because the tweezer potentials are not infinitely stiff, and thus optical forces from the emergent cavity field shift the atomic ensembles to positions away from their trap minima. Figure~\ref{Fig4}a shows a typical example of this effect:  After memory recall, several ensembles are observed at locations up to 4~$\mu$m away from the trap minima, which is 20\% of the tweezer waist. Moreover, the data indicate a spin-configuration dependence to the movement, because certain ensembles move differently when recalling one memory pattern versus another.   Reducing the transverse pump power down toward zero reverts the ensembles back to the tweezer trap centers, rendering this phenomenon a form of elastic rather than plastic response.  This elastic response acts to change the positions, and therefore the connectivity. Simulations in section IV.C~\cite{Supp} indicate that the motional shift deepens the energy well of the spin configurations through which the system evolves, which then enhances recall probability of the nearby memory. This can lead to the generation of new memory patterns that are self-consistently stable under both spin and positional dynamics, increasing memory capacity.

We experimentally verify that the $\mathbf{J}_1$ memory capacity increases when the polaronic deformations are made stronger. The laser power of each tweezer is lowered by a factor of four to yield a weaker, and therefore more elastic, trap. Figure~\ref{Fig3}a shows that this yields an almost two-fold increase in memory capacity for $\mathbf{J}_1$, resulting in 25(2) memories in a $n=16$ network, 7-times the Hopfield capacity.  We also measure an increased elastic response:  The positions shown in Fig.~\ref{Fig4}b are on average $\sim$38\% further away from the initial tweezer locations.  However, it does not saturate the basin volume bound quite as well.

In addition to enhancing capacity, simulations indicate that polaronic elasticity inhibits the effects of ``$J$-chaos" on recall and memory capacity; $J$-chaos refers to large changes in local minima induced by small changes to the $J_{ij}$~\cite{Bray1987cno,Zhu2016bpo}. In simulations where elasticity is fully removed, the 0.5-$\mu$m-scale experimental drift of tweezer positions reduces the $\mbf{J}_1$ memory capacity by 36\%; see section IV.C~\cite{Supp}. In contrast, simulations with the level of elasticity present in the experiments of Fig.~\ref{Fig4}b show no reduction of memory capacity due to position noise. 

\newpage

Future work could engineer a long-term form of plasticity that persists even without the pump field.  This could be achieved, e.g., by shaping optical potentials to have additional minima in learned locations. Plastic self-reinforcement of memories in this quantum-optical system would realize a natural form of learning while enjoying capacity enhancement and robustness to coupling errors.  Photonic network denoisers and amplifiers are two possible applications once intracavity networks are increased in size using larger tweezer arrays~\cite{Manetsch2024ata}.  Capacity-enhanced quantum associative memory might be possible using intracavity ensembles acting as effective spin-1/2 degrees of freedom~\cite{Marsh2024ear}. 

In summary, this work experimentally reveals a new paradigm of associative memory, with enhanced memory capacity that goes far beyond the Hopfield limit, by exploiting naturally occurring energy minima in a {\it glassy} energy landscape that the original Hopfield model must avoid.  Such glassy minima can nevertheless yield robust memories in our system, because they are stabilized not only by steepest rather than stochastic energy descent, but also by a form of polaronic elasticity induced by self-reinforcing spin-motion coupling.  This yields a quantum-optical realization of the short-term synaptic plasticity found in neurobiology.

We thank Ronen Kroeze, Giulia Socolof, and Deven Bowman for helpful discussions and experimental assistance. We are grateful for funding support from the Army Research Office (Grant~\#W911NF2210261).  Y.J.~acknowledges support from the Q-NEXT DOE National Quantum Information Science Research Center. J.K.~acknowledges support from EPSRC (Grant No.~EP/Z533713/1). B.M.~acknowledges funding from the Stanford QFARM Initiative. H.H.~acknowledges support from the Stanford Shoucheng Zhang Graduate Fellowship.  S.G.~thanks the Schmidt Science Polymath program for support.

%

%%%%%%%%%%%%%%%%%%%%%%%%%%%%%%%%%%%%%%%%% Supplement

\clearpage
\onecolumngrid

 \let\oldaddcontentsline\addcontentsline
        \renewcommand{\addcontentsline}[3]{}
    
\section*{Supplementary Information}

        \let\addcontentsline\oldaddcontentsline

\renewcommand{\theHfigure}{S\arabic{figure}}
\tableofcontents

\renewcommand{\theequation}{S\arabic{equation}}
\renewcommand{\thefigure}{S\arabic{figure}}
\setcounter{figure}{0}
\setcounter{equation}{0} 
\setcounter{figure}{0}
\setcounter{equation}{0} 

%%%%%%%%%%%%%%%%%%%%%%%%%%%%%%%%%%%%%%%%%%%%%%%%%%%%%%%%%%%%%%%%
\section{Experimental methods}
\label{sec:methods}

Networks of ultracold gases of $^{87}$Rb are prepared in a multimode optical cavity as described in Ref.~\cite{Marsh2025amc}, with adjustments as follows. An average of $M=4.2(3)\times 10^4$ atoms are trapped at each site of the network and have been evaporatively cooled below the critical temperature for Bose-Einstein condensation (BEC); we measure an average BEC fraction of 17\% with a 4\% variation across sites. While Bose-condensation aids in preparing low-entropy initial states of the neural network, it is not a requirement for this work. 

The atom numbers are controlled to balance the signal strength per site during readout, as described below. This results in an approximately 25\% variation in atom number between sites that does not change between experimental cycles. This variation does not degrade recall performance. Rather, it serves to roughly equalize the strength of the cavity-mediated interactions between ensembles by balancing the number of photons emitted by each spin ensemble.  

Optical tweezers are formed by crossed dipole traps, each with approximately a 20-$\mu$m waist and trap frequencies of $[\omega_x,\omega_y,\omega_z]=2\pi{\cdot}[326(5), 472(16), 332(9)]$~Hz. This yields an atomic density distribution $\rho(\mbf{r})$ with a $1/e$ radius of approximately 7~$\mu$m at each site. The tweezers form a rectilinear grid of $n_r$ rows and $n_c$ columns. The BECs are trapped at the $n=n_r\cdot n_c$ vertices of the grid to create networks of size $n$ in the midplane of the cavity. Different neural networks are realized by selecting random locations for the rows and columns of the grid, following the method described in Ref.~\cite{Marsh2025amc}. The minimum spacing between rows and columns is 40~$\mu$m, and the total spatial extent of the grid never exceeds 150~$\mu$m. The center of the grid deviates from the center of the cavity by up to 10~$\mu$m. The column locations for $\mbf{J}_1$ shown in Fig.~\ref{Fig4} of the main text are $[x_1,x_2,x_3,x_4]=[-79,-23,34,89]$~$\mu$m and the row locations are $[y_1,y_2,y_3,y_4]=[-94,-31,28,91]$~$\mu$m, both with respect to cavity center. 

A ``4/7" multimode optical cavity mediates atomic interactions to realize the $J_{ij}$ connections of the neural network. The cavity is formed by a pair of $R=1$~cm radius of curvature mirrors separated by a distance $L\approx 1.22$~cm, with a free spectral range (FSR) of $2\pi{\cdot}12.30010(6)$~GHz.  
The field decay rate reported in the main text is nearly constant for the modes participating in the 4/7 resonance~\cite{Kroeze2023hcu}. The cavity geometry approximately satisfies the $M/N$ multimode degeneracy condition $L/R=2\sin^2(M\pi/2N)$ for the irreducible fraction $M/N=4/7$~\cite{Guo2019eab,Marsh2025amc}. This means that the cavity hosts $N$ families of near-degenerate modes per FSR. Each family is identified by an integer $0\leq\eta<N$ and contains the set of all Hermite-Gaussian modes $\Xi_{lm}$ with constant $l+m\equiv\eta \pmod N$, up to a high-order mode cutoff~\cite{Kroeze2023hcu}. We choose the family with $\eta=0$ in this work for simplicity; the remaining mode families are far-detuned from the transverse pump frequency and do not contribute. The number $M$ controls the longitudinal character of the mode families; cavities with odd $M$ support standing-wave mode superpositions of arbitrary phase in the cavity midplane, while those superpositions at the midplane of even-$M$ cavities are restricted to being either $0$ or $\pi$ in phase. By employing an even-$M$ cavity in this work, we realize cavity-mediated interactions of Ising form~\cite{Marsh2025amc}, rather than the vector form realized in odd-$M$ cavities like confocal cavities~\cite{Kroeze2025dor}. The explicit form of the cavity-mediated interaction $J_{ij}$ is provided in Eq.~\eqref{eq:J}.

The atomic ensembles scatter photons from the $\lambda=780$-nm transverse pump into the cavity modes, realizing a multimode variant of the Hepp-Lieb-Dicke model~\cite{Hepp1973ots,Kirton2018itt,Mivehvar2021cqw}. The system lies in the large cooperativity limit of cavity QED with a single-atom, single-mode coupling strength $g_0=2\pi{\cdot}1.35$~MHz, excited-state decay rate $\Gamma=2\pi{\cdot}6.065$~MHz, and single-mode cooperativity $C=4.28$~\cite{Marsh2025amc}. Multimode enhancement in the dispersive coupling limit~\cite{Kroeze2023hcu} yields an estimated multimode cooperativity of $C_\text{mm}= 29$. The pump frequency is detuned by $\Delta_A\approx-2\pi{\cdot}97$~GHz from the D$_2$ transition of $^{87}$Rb and by $\Delta_C=-2\pi{\cdot}20$~MHz from the $\eta=0$ cavity resonance frequency. Superradiant scattering into the cavity occurs above a critical pump strength $\Omega_c$ given by $M\Omega_c^2g_0^2/\Delta_A^2=-2E_r(\Delta^2_C+\kappa^2)/(\Delta_C \lambda_\text{max})$, where $E_r=3.7$~kHz is the atomic recoil energy, $\lambda_\text{max}>0$ is the largest eigenvalue of the $J$ matrix~\cite{Marsh2024ear}, and small dispersive shifts have been disregarded. Threshold-less scattering occurs in the presence of the longitudinal pump fields. 

Self-organization of each BEC into one of two possible ``checkerboard" density wave states occurs concomitantly with superradiant scattering.   The density-wave states map onto an SU(2) collective spin degree of freedom $\hat{S}_i^{x/y/z}$ for each BEC in the network~\cite{Marsh2025amc}. Holographic imaging of light emitted from the cavity~\cite{Guo2019spa} provides individual spin-state readout of the $\ex{\hat{S}_i^x}$ spin components as well as the center-of-mass (CoM) positions $\mbf{r}_i$ for each BEC after processing of the cavity field, as shown in Fig.~\ref{Fig1}d and described in Ref.~\cite{Marsh2025amc}. Inhomogeneity in the magnitude of the measured spins $|\ex{\hat{S}_i^x}|$ naturally arises due to spatial variations in the multimode cavity coupling, as described in Sec.~\ref{sec:theory}. The atom numbers per site are tuned to minimize this effect, yielding measured $\ex{\hat{S}_i^x}$ that vary in magnitude by less than 10\% across sites, on average. Only the sign of $\ex{\hat{S}_i^x}$ is considered when judging whether a memory has been successfully recalled.

\begin{figure}[t]
    \centering
    \includegraphics[width=0.6\linewidth]{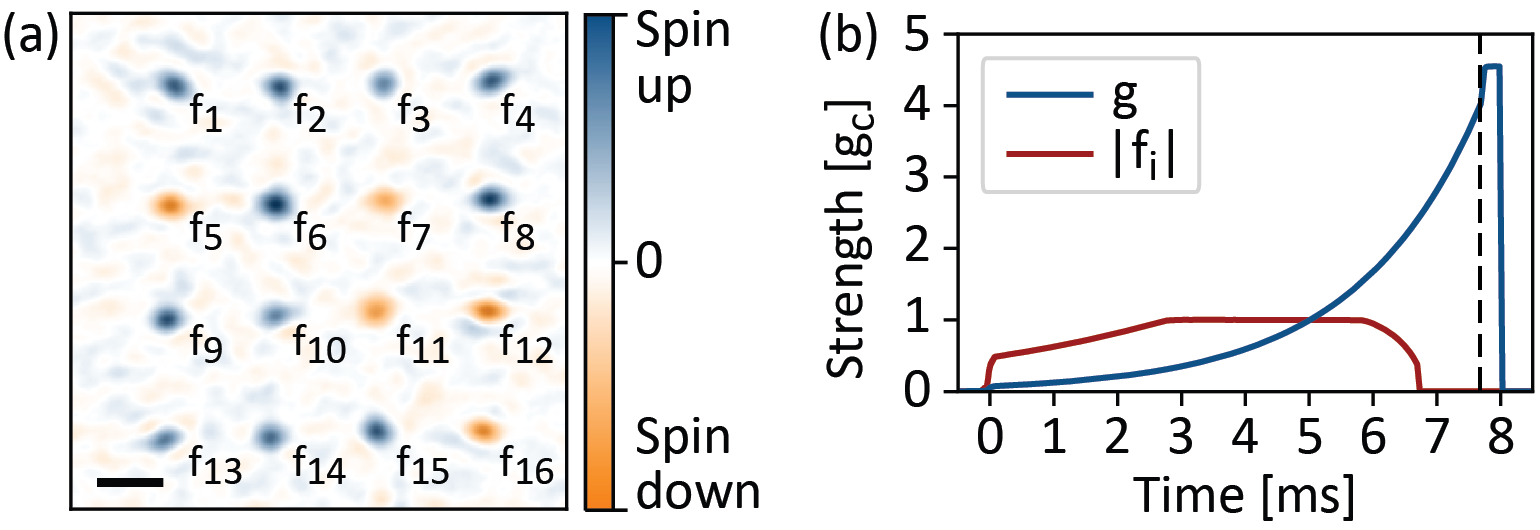}
    \caption{a) Example stimulus field imaged through the cavity in the absence of atoms. The image is normalized to its maximum amplitude and processed to remove nonlocal components of the cavity field~\cite{Marsh2025amc}. Each bright spot indicates a longitudinal field $f_i$ that couples to spin $\hat{S}_i^x$. The black bar indicates the waist size $w_0=34.8$~$\mu$m for the fundamental mode of the cavity. b) Transverse (blue) and longitudinal (red) pump schedule for recall experiments. The dashed line indicates where imaging begins for spin readout.}
    \label{fig:ramps}
\end{figure}

Longitudinal fields $f_i$ can stimulate arbitrary spin configurations and enable associative memory recall. To generate the fields, a portion of the transverse pump laser is split off, shaped by a digital micromirror device (DMD), and injected directly (longitudinally) into one of the cavity mirrors. The field is shaped in the Fourier plane by the DMD~\cite{Papageorge2016ctm,Guo2021aol} to yield an array of $n$ beams propagating in parallel to the cavity axis. Each individually targets an atomic ensemble at its CoM location $\mbf{r}_i$, as illustrated in Fig.~\ref{fig:ramps}a. Each beam is focused to within 1~$\mu$m of $\mbf{r}_i$ with a waist of 7.3~$\mu$m.  Their intensities and phases are fully tunable with respect to the transverse pump. The relative intensities are matched between sites to within 10\%. The longitudinal field strengths $|f_i|$ are proportional to the geometric mean of the longitudinal beam intensity and the transverse pump intensity, as described in Sec.~\ref{sec:theory}. The relative phases between the longitudinal beams and the transverse pump encode the spin state that stimulates the atoms: A spin-up (spin-down) stimulus corresponds to 0 ($\pi$) phase with $f_i>0$ ($f_i<0$). The phases are controlled to within ${<}0.3$ radians between different beams. However, the global phase between the longitudinal beams and the transverse pump is uncontrolled and exhibits a slow drift between experimental cycles. As shown in Sec.~\ref{sec:theory}, this introduces an amplitude factor $f_i\to\cos(\phi)f_i$ for the random ($i$-independent) angle offset $\phi$ arising in each experimental sequence. This does not affect the spin configuration being stimulated, but does affect the strength of the applied stimulus. Weak stimuli can inhibit associative memory recall, though we are able to mitigate this problem by increasing the strength of the longitudinal beams.  The simulations of Sec.~\ref{sec:sims} include all these noise effects in their estimates of memory capacity.

The ramp schedules for the longitudinal fields and transverse pump are optimized for associative memory recall fidelity and are presented in Fig.~\ref{fig:ramps}b. The transverse pump is ramped exponentially to $g=4g_c$, where $g_c=g_0\Omega_c/\Delta_A$ is the threshold coupling strength for the superradiant transition. This drives spin-flip dynamics away from the stimulated state and towards memory states. The pump is quenched at $7.7$~ms to $\sim4.5g_c$ for 300~$\mu$s to maximize photon flux for spin readout. No longitudinal fields are present during imaging. We find that the recall fidelity and basin size of memories are insensitive to small changes in this ramp schedule, such as making the entire sequence up to 20\% longer or shorter, or adjusting the precise time at which the longitudinal fields are turned off.

Each experimental trial yields a single output of the neural network, i.e., the spin states $\ex{\hat{S}_i^x}$. The $\ex{\hat{S}_i^x}$ are extracted through holographic imaging~\cite{Kroeze2018sso,Guo2019spa} of the cavity output over a 300~$\mu$s period as described in Ref.~\cite{Marsh2025amc} and summarized as follows: Each image is demodulated to produce a phase-sensitive image of the electric field in the cavity midplane before downsampling ${\times} 4$ in both directions and applying a fractional Fourier filter to remove noise. The images are then fit to a model involving the known cavity Green's function in Eq.~\eqref{eq:Geta} and processed to remove nonlocal fields. This leaves an image of the light scattered by each atomic ensemble, the phase and amplitude of which reveals each of their spin states $\ex{\hat{S}_i^x}$. For simplicity, we choose to normalize the spin states such that $\sum_{i}^n\ex{\hat{S}_i^x}^2=1$. 

Rarely, an atomic ensemble splits into two unequal components that scatter photons with opposite phase. This splitting can reduce the amplitude of $\ex{\hat{S}_i^x}$, which is the average over the atomic ensemble, but the amplitude remains above our noise floor in more than 99\% of the cases~\cite{Marsh2025amc} and thus minimally impacts our measurements. The above analysis is performed for all images shown in the main text. Additionally, the fit results provide the positions $\mbf{r}_i$ of the atomic ensembles in the cavity midplane. Slow feedback between trials keeps the array's CoM position $\mbf{r}_\text{CoM}=\sum_{i=1}^n\mbf{r}_i/n$ stabilized to within 0.5~$\mu$m of the intended target position. Experimental trials with a $\mbf{r}_\text{CoM}$ that deviates from the target by more than 1~$\mu$m are discarded, which occurs in approximately 8\% of trials for our $n=16$ neural networks~\cite{Marsh2025amc}.  

%%%%%%%%%%%%%%%%%%%%%%%%%%%%%%%%%%%%%%%%%%%%%%%%%%%%%%%%%%%%%%%%%%

\section{Associative memory methods}
\label{sec:AMmethods}
\begin{figure}[t]
    \centering
    \includegraphics[width=0.95\linewidth]{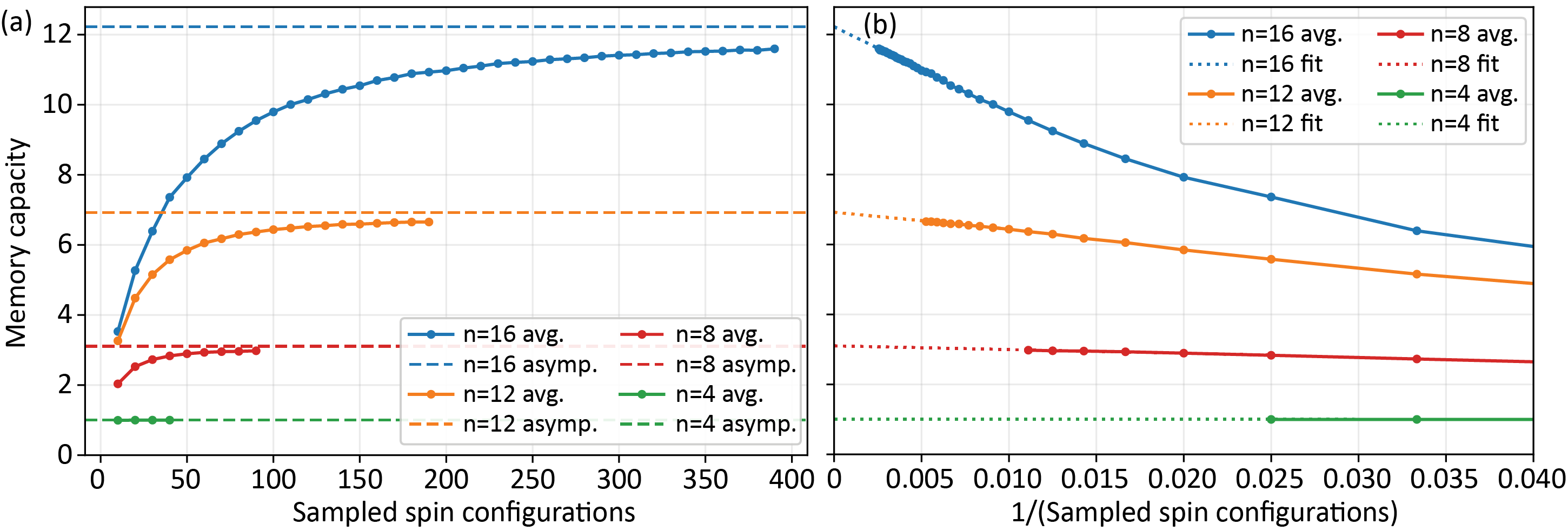}
    \caption{ Bootstrap estimation of the experimental memory capacity. a) Bootstrap samples are generated by resampling with replacement from the set of spin configurations found via random sampling. The number of memories present in each bootstrap sample is counted and averaged over 500 independent bootstrap samples for each sample size on the $x$ axis. The resulting curves of memory capacity versus sample size are averaged over $\mathbf{J}$ disorder realizations for each $n$. The asymptote of each curve, corresponding to the true, average memory capacity at each $n$, is estimated via a scaling analysis shown in panel (b). This analysis estimates that for $n=[16, 12, 8, 4]$ we find [95\%, 96\%, 96\%, 99\%] of the total number of memories. b) The same data as panel (a) are plotted versus the inverse of the number of samples. The $y$-intercepts of the curves thus correspond to the asymptotic memory capacities. These are estimated by linear extrapolation of the data to $x=0$.}
    \label{fig:RDMDshots}
\end{figure}

In this section, we describe how natural memories of the neural networks are found, grouped, and measured for their basin size. The memories are found by cataloging $\ex{\hat{S}_i^x}$ spin configurations over many trials with random inputs. The attractors of the spin dynamics are the memory states; thus, by allowing random input states to evolve, we generate a random sample of the memories. With enough random samples, all memories of the neural network can be found with high probability. We use [400, 200, 100, 50] samples per neural network for system sizes $n=[16, 12, 8, 4]$. A bootstrap analysis of the number of identified memories versus the number of samples is described in Fig.~\ref{fig:RDMDshots}. We estimate that for $n=[16, 12, 8, 4]$ we find, on average, $[95\%, 96\%, 96\%, 99\%]$ of the total number of memories with basin size ${\geq}1$ spin flip at the 50\% recall threshold level. It becomes increasingly challenging to find all memories as $n$ increases. Identifying all memories is by no means a requirement to operate the neural network, and we only do so here to study the scaling of the memory capacity. Indeed, the difficulty of finding all memories at higher $n$ highlights the significantly enhanced memory capacity over the Hopfield model.  

\begin{figure*}[t]
    \centering
    \includegraphics[width=\linewidth]{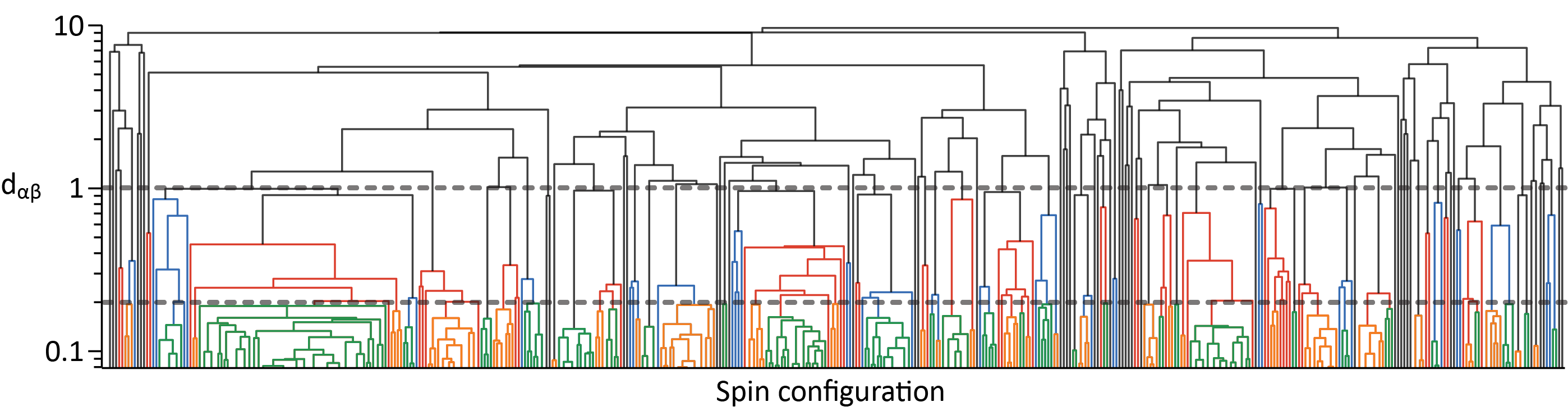}
    \caption{Hierarchical clustering of the randomly sampled spin configurations for $\mbf{J}_1$. The dendrogram shows the formation of spin configuration clusters at different levels of the overlap distance $d_{\alpha\beta}$ given by Eq.~\eqref{eq:olapDist}. The upper dashed line at $d_{\alpha\beta}=1$ cuts the dendrogram at the single-spin-flip level to define memory candidates. Candidates containing a cluster of spin configurations are color-coded  red or blue in an alternating fashion for visual clarity; candidates corresponding to single spin configurations remain as black lines below the $d_{\alpha\beta}=1$ threshold. The lower dashed line shows the two-sigma confidence level above the noise floor. Clusters that form below this level contain all the same spin configuration to within experimental uncertainty.  They are recolored green or orange for clarity. See text for details.
    }
    \label{fig:dendro}
\end{figure*}

The randomly sampled spin configurations for each neural network are organized into a set of memory candidates to be tested. This is accomplished using a hierarchical clustering~\cite{Bar-Joseph2001fol} of the spin states by their overlap distance,
\begin{equation}\label{eq:olapDist}
    d_{\alpha\beta}=\frac{n}{2}\bigg(1-\Big|\sum_{i=1}^n\ex{\hat{S}_{i\alpha}^x}\ex{\hat{S}_{i\beta}^x}\Big|\bigg),
\end{equation}
where $\alpha$ and $\beta$ label the independently sampled spin configurations. The prefactor $n/2$ scales the distance such that $d_{\alpha\beta}$ is equal to the Hamming distance for spin states with uniform amplitudes. Thus, a distance $d_{\alpha\beta}=1$ approximately corresponds to one complete spin flip between spin configurations $\alpha$ and $\beta$. The hierarchical clustering outputs a grouping of the spin configurations that can be visualized as a tree of related states, or more precisely as a dendrogram. 

The dendrogram for $\mbf{J}_1$ is shown in Fig.~\ref{fig:dendro}, relating to data shown in Fig.~\ref{Fig1} and Fig.~\ref{Fig2} of the main text. The tree branches into many clusters of similar spin configurations; each vertical line in the dendrogram is one such cluster, and the horizontal lines at branch points show the average $d_{\alpha\beta}$ between states in the cluster. The tree further branches into sub-clusters of more closely related states until reaching the leaves at the bottom of the plot, each corresponding to a single spin configuration. This branching arises due to the ultrametric structure of the spin glass we create~\cite{Marsh2025amc}.  Memory candidates are formed by cutting the tree at the upper dashed line where $d_{\alpha\beta}=1$; each intersection of the upper dashed line with the tree yields a cluster of spin configurations that, on average, differ by less than one complete spin flip within the cluster. We therefore consider each of these clusters to be a memory candidate, containing a few closely related spin configurations, and color code them alternately red or blue for visual clarity. Spin configurations that are not clustered with any others below the $d_{\alpha\beta}=1$ cutoff are also considered as candidates and remain as black lines below both cutoffs. The lower dashed line at $d_{\alpha\beta}=0.21$ is the two-sigma confidence level above the noise floor of our measurements. Clusters that form below this level contain spin configurations that are the same to within experimental uncertainty; these clusters are colored green and orange in an alternating fashion. This grouping procedure for memory candidates is found to enhance the stability of associative memory recall over long durations, combating technical noise and experimental sources of drift, while retaining a unique identity for each memory. 

Memory candidates are tested for their recall performance to either be deemed a valid memory or ignored. Only the candidates that exhibit a basin size greater than or equal to one spin flip at the 50\% recall level are accepted as memories and contribute to the memory capacity of the neural network. See Sec.~\ref{sec:sims} and Sec.~\ref{sec:extendedData} for a discussion of the dependence of memory capacity  on the 50\% recall threshold. Each candidate is first tested by measuring its recall probability over 30 trials with a stimulus spin configuration that exactly matches the sign signature of the candidate memory; in other words, with zero errors in the input. In the case that the candidate cluster contains a small number of different sign signatures, the one that was found most commonly during random sampling is used as the reference. We find that if the zero-error recall probability is less than 75\%, the likelihood of the basin size being greater than or equal to 1 spin flip is very low. Thus, candidates with zero-error recall probability less than 75\% are ignored without further testing. Passing candidates are then measured for their basin size under non-zero error in the stimulus. We note that the grouping procedure for memory candidates could inflate the non-zero-error recall probabilities. This is because applying errors to the reference spin configuration could lead to a stimulus that still matches with another spin configuration in the candidate cluster. However, the effect diminishes with increasing $n$ and occurs in less than 10\% of the cases at $n=16$ for stimuli with single-spin-flip errors. The effect is negligible for stimuli with two or more errors.   

Basin sizes are measured over 30 additional recall trials with a variable number of errors. The case of $\mbf{J}_1$ is the exception:  We measured full recall curves more thoroughly, conducting 30 trials per error level. But for all other $\mbf{J}$s, the basin size is estimated in an iterative approach as follows. The number of errors in the first of the 30 additional recall trials is chosen randomly between 1 and $n/2$. The result is recorded as either a successful or failed recall of the memory at the randomly chosen error level. This data point, combined with the already measured zero-error recall probability, yields a sparse estimate of the memory's recall curve. The recall curve is fit to produce an estimate of the basin size after one recall trial, $b_1$. Additional recall trials yield a more accurate estimate of the recall curve and thus a more accurate estimate of the basin size. After the first trial, the number of stimulus errors used in the second trial is again randomly sampled between 1 and $n/2$, but with a bias toward $b_1$. 
Biasing the number of spin flip errors toward the estimated basin size provides the most information about the true basin size. This is implemented by sampling the number of spin flip errors for trial $i+1$ from the probability distribution $P_{i+1}(e)\propto1/[(e-b_i)^2+1.5^2]$, where $e$ is the number of errors and $b_i$ is the estimated basin size after trial $i$. We use a distribution width of 1.5 spin flips. The basin size estimate is not sensitive to the precise functional form. This ad-hoc distribution was found to improve the accuracy of the final estimated basin size as compared to uniform random sampling.  The second recall trial yields a more accurate estimate of the recall curve and more accurate basin size estimate, $b_2$. This procedure continues over 30 total trials, yielding a final estimate $b_{30}$ that is the most accurate. 

Basin sizes are extracted through a least-squares fit of the recall curve to a functional form $a_1[1-\tanh(a_2e-a_3)]$, where $a_{1/2/3}$ are the fit parameters. The intersection of the fitted curve with the 50\% recall probability threshold defines the basin size. We find that fitting the recall curves to estimate the basin size is less sensitive to statistical fluctuations than directly interpolating the data to find the 50\% crossing point. Though chosen in an ad hoc manner, in Sec.~\ref{sec:extendedData} we show that the $\tanh$ functional form fits well to the observed recall curves.  Moreover, the estimated basin sizes are insensitive to the precise functional form. The measurement uncertainty in the basin size is estimated via a bootstrap analysis: The 60 recall trials (30 without stimulus errors and 30 with stimulus errors) are resampled with replacement in 100 bootstrap samples, resulting in 100 bootstrapped recall curves. The basin size for each recall curve is estimated through the same $\tanh$ fit. The standard deviation in the basin sizes provides an estimate of the uncertainty, which is typically around 0.3 spin flips. 

The memory capacity for each neural network is determined by counting the number of memories with a basin size greater than or equal to one spin flip. Uncertainty in the basin sizes leads to uncertainty in the memory capacity. This uncertainty in the capacity is estimated through an additional bootstrap analysis. For each neural network, over 1,000 bootstrap samples of each basin size are generated by adding Gaussian-sampled noise to the basin estimate. The noise level is set by the uncertainty in the basin size estimate. Each bootstrap sample contains a noisy estimate of the basin size for each candidate and leads to a single bootstrap estimate of the memory capacity. The mean of the bootstrap distribution is the estimated memory capacity of the neural network that we quote in the main text. The standard deviation of the bootstrap distribution provides the estimated uncertainty in the memory capacity for that neural network. This uncertainty is typically on the scale of $\pm1$ memory. 

In addition to measuring basin size, a basin volume can also be measured for each memory. The volume of a memory's basin of attraction is the number of stimulus spin configurations that dynamically evolve to the memory. The volume can be generalized to accommodate stochastic dynamics as follows. Each of the $2^n$ stimulus spin configurations $\mbf{s}=(\pm1,\cdots,\pm1)$ have some probability to evolve to a given memory $i$. If we denote this probability as $p_i(\mbf{s})$, then the basin volume for memory $i$ is $V_i=\sum_\mbf{s}p_i(\mbf{s})$, where the sum is taken over all $2^n$ spin configurations. Conveniently, the basin volume may be estimated from the random sampling of spin configurations that is initially performed to identify memory candidates. After determining the memories, an estimate of the basin volume is given by $2^nN_i/N_\text{total}$, where $N_i$ is the number of times memory $i$ is encountered in the random sampling of states and $N_\text{total}$ is the total number of random samples. Uncertainty in this estimation is computed through bootstrap resampling of the randomly sampled spin configurations. The average basin size at $n=16$ is approximately 5,600 spin configurations, with an average uncertainty of 10\%.

%%%%%%%%%%%%%%%%%%%%%%%%%%%%%%%%%%%%%%%%%%%%%%%%%%%%%%%%%%%%%%%%%%

\section{Theoretical description}\label{sec:theory}

This section provides the theoretical framework describing the quantum-optical neural networks we create. We first provide a quantum-optical model that explicitly includes the quantum fields of the multimode cavity. An effective spin model is then derived in which the photonic degrees of freedom are eliminated in the limit of dispersive cavity coupling. We then simplify the description further in a semiclassical model that serves as the basis for the subsequent numerical studies in Sec.~\ref{sec:sims}.

\subsection{Quantum-optical model}

Each associative memory neural network that we create is formed from a network of $n$ atomic ensembles trapped within a multimode optical cavity. The system is driven-dissipative: The ensembles scatter photons from a transverse pump into the cavity, where the photons may scatter from another atomic ensemble, producing a cavity-mediated spin-spin interaction, or dissipate out of the cavity. Each ensemble forms one of the spins in the network, where the collective pseudospin states are formed from motional states of the atomic gas~\cite{Marsh2025amc}. The motional state in the externally applied trap, without transverse pumping, defines the normal state $\psi_0$ with $\ex{\hat{S}_i^z}=-S$. The atomic wavefunction $\psi_c\propto \cos(k_rz)\cos(k_r x)$ defines the pseudospin state with $\ex{\hat{S}_i^z}=S$, where $k_r=2\pi/\lambda$. This state arises due to the potential formed from the interference of the transverse pump with the emergent cavity field. Two ``checkerboard" density wave states proportional to $\psi_0+\psi_c$ with $\ex{\hat{S}_i^x}=S$ and $\psi_0-\psi_c$ with $\ex{\hat{S}_i^x}=-S$ define what we refer to as the spin up and spin down states, respectively. Each atomic ensemble is thus described by a collective SU(2) spin operator $\hat{S}_i^{x/y/z}$ with total spin $S=M/2$~\cite{Marsh2025amc}. The quantum-optical description of the system is provided by a multimode, multi-atomic ensemble generalization of the Hepp-Lieb-Dicke model~\cite{Kirton2018itt,Marsh2025amc}, given by
\begin{equation}
    \frac{\hat{H}_\text{Total}}{\hbar} = \sum\limits_\mu\Big[ -\Delta_\mu \hat{a}_\mu^\dagger \hat{a}_\mu + f_\mu( e^{i\phi_\mu}\hat{a}_\mu  + e^{-i\phi_\mu}\hat{a}_\mu^\dag) \Big] +\omega_z\sum_{i=1}^n \hat{S}_i^z + \sum_\mu\sum_{i=1}^ng_{\mu}(\mbf{r}_i) \hat{S}_i^x(\hat{a}_\mu+\hat{a}_\mu^\dag).
\end{equation}
The first sum describes the multimode cavity under longitudinal pumping. It includes all the Hermite-Gauss modes $\Xi_\mu$ within one of the near-degenerate mode families of the $M/N=4/7$ cavity. The mode labels $\mu=(l,m)$ index the number of nodes in the mode profile in the two transverse directions. A mode family contains those $\Xi_\mu$ for which $l+m \equiv\eta \pmod N$ for a fixed integer $\eta\in[0,N-1]$. We employ an $\eta=0$ cavity in this work; see Sec.~\ref{sec:methods} for details about the cavity. The operators $\hat{a}_\mu$ satisfy canonical bosonic commutation relations $[\hat{a}_\mu,\hat{a}_\nu]=[\hat{a}^\dag_\mu,\hat{a}^\dag_\nu]=0$ and $[\hat{a}_\mu,\hat{a}^\dag_\nu]=\delta_{\mu\nu}$. The transverse pump is red-detuned from each mode by $\Delta_\mu=\omega_P-\omega_\mu<0$. We denote the detuning from the fundamental Hermite-Gauss mode $\Xi_{0,0}$ by $\Delta_C=\Delta_{0,0}$. A longitudinal pump shaped by a DMD coherently drives the cavity modes with strengths $f_\mu$ and phases $\phi_\mu$. The sum over $\hat{S}_i^z$ operators describes the $\hbar\omega_z\approx 2E_r$ energy cost for the formation of the atomic density wave state, where $E_r \approx h\cdot3.8$~kHz is the atomic recoil energy and $h$ is Planck's constant. The final sum describes the light-matter interaction between each atomic ensemble and cavity mode. Given the CoM position $\mbf{r}_i$ of an atomic ensemble in the cavity midplane, the position-dependent coupling strengths are given by
\begin{equation}
    g_{\mu}(\mbf{r}) = \cos\theta_\mu \frac{g_0\Omega}{\Delta_A}\int d\mbf{r}'\rho(\mbf{r}'-\mbf{r})\Xi_\mu(\mbf{r}'),
\end{equation}
where $g_0=2\pi\cdot 1.35$~MHz is the single-atom, single-photon coupling strength, $\Omega$ is the Rabi frequency of the transverse pump, and $\Delta_A\approx-2\pi\cdot97$~GHz is the excited-state detuning of the transverse pump. The terms $\Xi_\mu(\mbf{r})$ are spatial mode profiles at the midplane of the cavity and $\rho(\mbf{r})$ is the overall envelope of the atomic density profile. $\rho(\mbf{r})$ is well approximated by an isotropic 2D Gaussian distribution with a standard deviation of approximately 5.2~$\mu$m. The mode phases $\theta_\mu$ simplify to $(l+m)2\pi/7-(1+Q_0)\pi/2$ in the midplane of the $\eta=0$ cavity that we consider, where $Q_0$ is the longitudinal mode number of the $\Xi_{0,0}$ fundamental mode; see Ref.~\cite{Marsh2025amc} for the general form of $\theta_\mu$. The atomic density distribution in the cavity midplane $\rho(\mbf{r})$ is integrated against the Hermite-Gaussian mode profile $\Xi_\mu(\mbf{r})$ to yield the coupling strength. 

Dissipation of the cavity modes is captured by the Lindblad master equation 
\begin{equation}\label{eq:master}
    \frac{d}{dt}\rho=-\frac{i}{\hbar}[\hat{H}_\mathrm{Total},\rho]+\kappa\sum_\mu\big(2\hat{a}_\mu\rho\hat{a}_\mu^\dag - \{\hat{a}_\mu^\dag\hat{a}_\mu,\rho\} \big),
\end{equation}
where $\rho$ is the density matrix of the full system. The field decay rate $\kappa=2\pi{\cdot}140$~kHz is approximately constant among the cavity modes~\cite{Kollar2015aac}.

\subsection{Effective spin model}

We now derive a simplified theory in which the photonic degrees of freedom are adiabatically eliminated to yield an effective transverse field Ising model. The position dependence of the spins is retained to allow a description of spin-position coupling in the effective Hamiltonian. We identify this spin-position coupling as the reason for the polaronic deformation of the spin positions noted in the main text. Moreover, we show that the effective model we derive is able to reproduce the enhanced memory capacity due to polaronic deformation. The primary result of this section is to provide an effective theory that plausibly explains the mechanism behind the enhanced memory capacity due to polaronic deformation. We note that more advanced theoretical treatments are necessary to derive an effective model that captures the steepest descent dynamics of the spins, which we performed in Ref.~\cite{Marsh2021eam}, and to derive an effective model that fully captures the form of the dissipation channels, which we performed in Ref.~\cite{Marsh2024ear}. Deriving a theory that describes all aspects of the spin dynamics in a simplified theory would go well beyond the scope of the current experimental study, but is something that future theoretical work should address. Rather, we focus here on providing a simplified theory that captures the polaronic enhancement to memory capacity.   

The photonic degrees of freedom are eliminated by inserting the steady-state expressions for the cavity field into the equations of motion for the spin degrees of freedom. This approach captures the effective Hamiltonian seen by the atoms, as discussed below, but does not give a complete description of dissipation due to the retardation of cavity-mediated interactions. The resulting effective model is accurate in the limit of dispersive cavity coupling, $|\Delta_C|\gg\omega_z,\kappa,|g_\mu(\mbf{r})|$, and the spread of the near-degenerate cavity modes, which are all approximately satisfied in our experimental parameter regime with $\Delta_C\approx2\pi\cdot20$~MHz, $\omega_z\approx2\pi\cdot7.5$~kHz, and $|g_\mu(\mbf{r})|$ on the kilohertz scale. The steady-state expressions for the cavity field operators under the Lindblad master equation presented in Eq.~\ref{eq:master} are given by
\begin{equation}\label{eq:modeExp}
    \hat{a}_\mu = \frac{f_\mu e^{-i\phi_\mu}}{\Delta_\mu+i\kappa}+\sum_{i=1}^n \frac{g_{\mu}(\mbf{r}_i)\hat{S}_i^x}{\Delta_\mu+i\kappa}.
\end{equation}
The steady-state expression above leads to the following Heisenberg equations of motion for the spins after adiabatically eliminating the cavity modes,
\begin{equation}\label{eq:EOMAtomOnly}
    \begin{split}
        \frac{d}{dt}\hat{S}_i^x &= - \omega_z \hat{S}_i^y, \\
        \frac{d}{dt}\hat{S}_i^y &=   \omega_z \hat{S}_i^x - f(\mbf{r}_i)\hat{S}_i^z -g\sum_{j=1}^n J(\mbf{r}_i,\mbf{r}_j)\big(\hat{S}_i^z \hat{S}_j^x + \hat{S}_j^x\hat{S}_i^z  \big), \\
        \frac{d}{dt}\hat{S}_i^z &=   f(\mbf{r}_i)\hat{S}_i^y +g\sum_{j=1}^n J(\mbf{r}_i,\mbf{r}_j)\big(\hat{S}_i^y \hat{S}_j^x + \hat{S}_j^x\hat{S}_i^y  \big),
    \end{split}
\end{equation}
where $g=g_0^2\Omega^2/(\Delta_A^2|\Delta_C|)$. The terms $J(\mbf{r},\mbf{r}')$ and $f(\mbf{r})$ are position-dependent Ising couplings and longitudinal fields, respectively, and will be described below. The equations of motion above are generated by the following transverse field Ising model,
\begin{equation}\label{eq:HatomOnly}
    \frac{\hat{H}}{\hbar} = \omega_z \sum_{i=1}^n \hat{S}_i^z -g\sum_{ij=1}^n J(\mbf{r}_i,\mbf{r}_j)\hat{S}_i^x\hat{S}_j^x - \sum_{i=1}^n f(\mbf{r}_i) \hat{S}_i^x.
\end{equation}

Equation~\eqref{eq:HatomOnly} above defines the effective spin-position coupling model. We now describe the two position-dependent terms. The first is the Ising coupling matrix, given by
\begin{equation}\label{eq:J}
\begin{split}
    J(\mbf{r}_i,\mbf{r}_j) &= \Delta_C^2\sum_\mu\int d\mbf{r} d\mbf{r}' \rho(\mbf{r}-\mbf{r}_i)\rho(\mbf{r}'-\mbf{r}_j) \frac{\Xi_\mu(\mbf{r})\Xi_\mu(\mbf{r}')}{\Delta_\mu^2+\kappa^2} \\
    &= \frac{\Delta_C^2}{\Delta_C^2+\kappa^2}\int d\mbf{r} d\mbf{r}' \rho(\mbf{r}-\mbf{r}_i)\rho(\mbf{r}'-\mbf{r}_j) G^\eta(\mbf{r},\mbf{r}'),
\end{split}
\end{equation}
where $G^\eta(\mbf{r},\mbf{r}')$ is the cavity Green's function. This is given by the following expression in the limit of ideal mode degeneracy~\cite{Marsh2025amc},
\begin{equation}\label{eq:Geta}
\begin{split}
    G^\eta\big(\mbf{r},\mbf{r}'\big) = \frac{1}{7}\delta\left(\frac{\mbf{r}-\mbf{r}'}{w_0}\right) +  \sum_{k=1}^3  \frac{1}{7\pi\sin(2k\pi/7)} \sin\left[ (1+\eta)\frac{2k\pi}{7} + \frac{\mbf{r}^2 + {\mbf{r}'}^2}{\tan(2k\pi/7)w_0^2} - \frac{2\mbf{r}\cdot\mbf{r}'}{\sin(2k\pi/7)w_0^2}\right].
\end{split}
\end{equation}
We obtain an explicit form of $J(\mbf{r}_i,\mbf{r}_j)$ using Gaussian atomic densities $\rho(\mbf{r})=\exp[-\mbf{r}^2/(2\sigma_A^2)]/(2\pi\sigma_A^2)$. The experimental atomic densities may be closely approximated using $\sigma_A=5.2$~$\mu$m. $J(\mbf{r}_i,\mbf{r}_j)$ can then be evaluated as~\cite{Kroeze2023hcu}
\begin{equation}\label{eq:Jexplicit}
    J(\mbf{r}_i,\mbf{r}_j) = \frac{\Delta_C^2}{\Delta_C^2+\kappa^2}\left(\frac{1}{7}G'(\mbf{r}_i,\mbf{r}_j,1)+ \frac{2}{7}\sum_{k=1}^3\Re\left[ e^{-\eta 2\pi ik/7}G'\big(\mbf{r}_i,\mbf{r}_j,e^{2\pi i k/7}\big)\right]\right),
\end{equation}
where $G'(\mbf{r},\mbf{r}',t)$ is a modified Mehler kernel given by
\begin{equation}
    G'(\mbf{r},\mbf{r}',t) = \frac{(1+\gamma)^2}{4(1-\gamma^2t^2)}\exp\left[ -\frac{(1+\gamma)}{2(1-\gamma^2t^2)}\left((1+\gamma t^2)\frac{(\mbf{r}^2+\mbf{r}'^2)}{w_0^2}-2(1+\gamma)t\frac{\mbf{r}\cdot\mbf{r}'}{w_0^2} \right) \right].
\end{equation}
Above, $\gamma = (1-2\sigma_A^2/w_0^2)/(1+2\sigma_A^2/w_0^2)$ incorporates the finite size of the atomic density.

The second position-dependent term is the function $f(\mbf{r})$. This is the total intracavity field generated by the longitudinal pump integrated against the atomic density, given by
\begin{equation}
    f(\mbf{r}) = 2\frac{g_0\Omega}{\Delta_A}\sum_\mu \frac{f_\mu  \cos\big[\phi_\mu+\atan(\kappa/\Delta_\mu)\big]}{\sqrt{\Delta_\mu^2+\kappa^2}}\int d\mbf{r}' \rho(\mbf{r}'-\mbf{r})\Xi_\mu(\mbf{r}').
\end{equation}
The angle $\atan(\kappa/\Delta_\mu)$ is on the order of $\pi/400$ in our parameter regime and is negligible. We now evaluate $f(\mbf{r})$ for the form of the longitudinal pump that we use in recall experiments: A sum of focused Gaussian beams at each position $\mbf{r}_i$ with a phase corresponding to a desired stimulus spin configuration $s_i=\pm1$. Experimental examples of $f(\mbf{r})$, processed to remove background nonlocal fields~\cite{Marsh2025amc}, are shown in Fig.~\ref{Fig1}b, Fig.~\ref{Fig2}a, and Fig.~\ref{fig:ramps}a. The input field $E_\text{in}(\mbf{r})$ from the longitudinal pump is more explicitly modeled as
\begin{equation}
    E_\text{in}(\mbf{r}) = \frac{f}{2\pi w^2}\sum_{i=1}^ns_i\exp\left[-\frac{(\mbf{r}-\mbf{r}^0_i)^2}{w^2}\right],
\end{equation}
where $\mbf{r}_i^0$ are the target positions in the cavity midplane, $f$ is an overall coupling strength and $w$ is the waist of Gaussian input beams. The coupling strengths $f_\mu$ are then given by the overlap integral with the cavity modes as $f_\mu=\int d\mbf{r} E_\text{in}(\mbf{r})\Xi_\mu(\mbf{r})$, and the mode phases are given by a constant $\phi_\mu=\phi$. The experimentally realized waist of the longitudinal pumping beams, $w=7.3$~$\mu$m, closely matches the shape of the Gaussian atomic distribution with $\sigma_A\approx 5.2$~$\mu$m: Writing $E_\text{in}(\mbf{r})$ in the form standard Gaussian distributions with standard deviation $\sigma=w/\sqrt{2}$ yields $\sigma\approx5.16$~$\mu$m, closely matching $\sigma_A$. In this case, $f(\mbf{r})$ can be well approximated by
\begin{equation}\label{eq:f_simplified}
\begin{split}
    f(\mbf{r})&=\frac{\cos(\phi)g_0\Omega f}{\Delta_A\sqrt{\Delta_C^2+\kappa^2}}\sum_\mu\sum_{i=1}^n s_i\int d\mbf{r}'d\mbf{r}'' \rho(\mbf{r}'-\mbf{r})\rho(\mbf{r}''-\mbf{r}^0_i)\Xi_\mu(\mbf{r}')\Xi_\mu(\mbf{r}'') \\
    &= \frac{\cos(\phi)g_0\Omega f\sqrt{\Delta_C^2+\kappa^2}}{\Delta_A\Delta_C^2}\sum_{i=1}^n s_i J(\mbf{r},\mbf{r}^0_i),
\end{split}
\end{equation}
where $J(\mbf{r},\mbf{r}')$ is the same function that describes the Ising coupling. In writing Eq.~\eqref{eq:f_simplified}, we also make the approximation of ideal mode degeneracy, $\Delta_\mu=\Delta_C$ for all $\mu$. This approximation becomes accurate in the dispersive coupling regime in which we operate.

\subsection{Semiclassical model}\label{sec:semiclassical}

A semiclassical description of the spin system becomes accurate far above the superradiant threshold when the cavity field enters a coherent state. The resulting model is a simplification that, nevertheless, captures the ability to describe associative memory recall, spin-position coupling, and enhanced memory capacity using numerically tractable equations of motion. The semiclassical model is derived through mean-field decoupling of operator products in the Hamiltonian and equations of motion for the effective spin model. Specifically, the semiclassical energy $E$ is derived by taking the expectation value of Eq.~\eqref{eq:HatomOnly}, factorizing product terms, and explicitly including the trap potentials as a function of the CoM coordinates of the atomic ensembles,
\begin{equation}\label{eq:Esemiclassical}
    E = \hbar\omega_z \sum_{i=1}^n \ex{\hat{S}_i^z} -\hbar g\sum_{ij=1}^n J(\mbf{r}_i,\mbf{r}_j)\ex{\hat{S}_i^x}\ex{\hat{S}_j^x} - \hbar\sum_{i=1}^n f(\mbf{r}_i) \ex{\hat{S}_i^x}+\frac{1}{2}E_\text{trap}\sum_{i=1}^n\left(\frac{\mbf{r}_i-\mbf{r}_i^0}{w_t}\right)^2,
\end{equation}
where $w_t\approx20$~$\mu$m is the waist of the external trap potential and $E_\text{trap}$ sets the energy scale in the harmonic approximation. Equations of motion for the spin degrees of freedom are derived from Eq.~\eqref{eq:EOMAtomOnly} by taking expectation values and factorizing operator products. We approximate the positional dynamics by damped classical dynamics in the energy landscape. This corresponds to the classical equation of motion 
\begin{equation}
    m\frac{d^2\mbf{r}}{dt^2}=-\frac{dE}{d\mbf{r}}-c_d\frac{d\mbf{r}}{dt},
\end{equation}
where $m$ is the atomic mass and $c_d$ is a damping coefficient. A nonzero $c_d$ can arise from, e.g., damping of collective oscillations in the BEC~\cite{Pethick2002}. When the optical forces $dE/d\mbf{r}$ are not too strong, the system is overdamped and the inertial term $md^2\mbf{r}/dt^2$ becomes negligible. In this limit, the equation of motion reduces to $d\mbf{r}/dt=-(1/c_d)dE/d\mbf{r}$, corresponding to a gradient descent in the energy landscape.   The combined semiclassical equations of motion for the spin and positional degrees of freedom are then given by 
\begin{equation}\label{eq:EOMsemiclassical}
    \begin{split}
        \frac{d}{dt}\ex{\hat{S}_i^x} &= - \omega_z \ex{\hat{S}_i^y}, \\
        \frac{d}{dt}\ex{\hat{S}_i^y} &=   \omega_z \ex{\hat{S}_i^x} - f(\mbf{r}_i)\ex{\hat{S}_i^z} -2g\sum_{j=1}^n J(\mbf{r}_i,\mbf{r}_j)\ex{\hat{S}_i^z}\ex{\hat{S}_j^x}, \\
        \frac{d}{dt}\ex{\hat{S}_i^z} &=   f(\mbf{r}_i)\ex{\hat{S}_i^y} +2g\sum_{j=1}^n J(\mbf{r}_i,\mbf{r}_j)\ex{\hat{S}_i^y}\ex{ \hat{S}_j^x}, \\
        \frac{d}{dt}\mbf{r}_i &= -\frac{1}{c_d}\frac{dE}{d\mbf{r}_i}.
    \end{split}
\end{equation}
 The above model serves as the backbone for the simulations we perform in Sec.~\ref{sec:semiclassicalSims}, where we investigate numerical solutions to the above equations of motion.

%%%%%%%%%%%%%%%%%%%%%%%%%%%%%%%%%%%%%%%%%%%%%%%%%%%%%%%%%%%%%%%%%%

\section{Numerical simulations}\label{sec:sims}

In this section, we perform numerical simulations of associative memory in the Hopfield model, SK model spin glass, and in the quantum-optical neural networks. We first benchmark the memory capacity of the Hopfield and SK models using the same memory criterion applied to our experimental data. We then model associative memory in the quantum-optical neural networks to demonstrate the enhanced capacity arising from polaronic deformations and to estimate the effect of experimental imperfections. 

\subsection{Hopfield model memory capacity}

In this section, we numerically determine the memory capacity of the Hopfield model subject to the same criterion that we apply to the determination of capacity in our experimental neural networks: a minimum basin size of one spin flip at the 50\% recall level.  The Hopfield model is known to have a memory capacity of approximately $0.138\cdot n$ in the large-$n$ limit~\cite{Amit1985sin}. However, this definition has certain issues: It permits small but extensive deviations from the intended memory patterns~\cite{Amit1985sin}, does not accurately model small-$n$ effects, and does not enforce a basin size of at least one spin flip. Thus, we numerically compute the Hopfield capacity to allow for direct comparison to our experimentally measured memory capacities presented in the main text.

We numerically determine the Hopfield memory capacity as follows. For each $n$, we optimize the number of intended patterns $P$ to maximize the total number of stored memories. The intended pattern vectors $\xi^p$ are stored using the standard Hebbian learning rule $J_{ij} = \sum_{p = 1}^P \xi_i^p \xi_j^p$~\cite{Hopfield1982nna}. This learning rule successfully stores patterns with high probability for small $P/N$, but as $P$ increases, the probability of successfully storing a pattern decreases; this tradeoff results in an optimal $P$ that stores the most patterns, on average. We determine the optimal $P$ by numerically checking the memory capacity in 10,000 trials with randomly chosen patterns for each value of $P$. A pattern is considered to be successfully stored if it exhibits greater than 50\% recall probability under MH dynamics with randomized single-spin-flip errors; this finds all basins of size greater than or equal to one spin flip.

\begin{figure}[t]
    \centering
    \includegraphics[width=\linewidth]{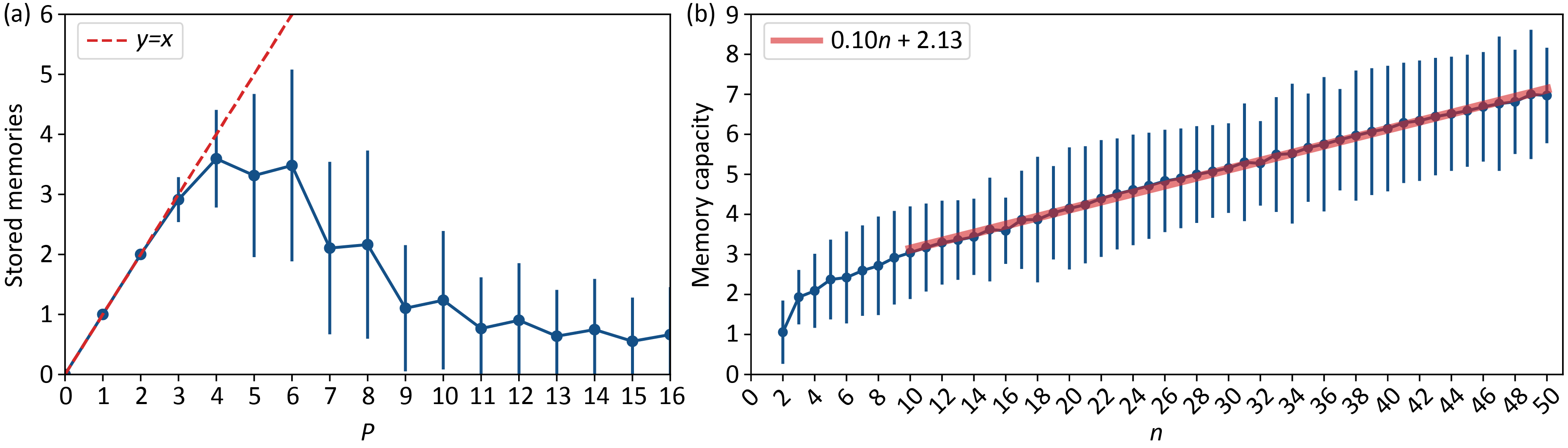}
    \caption{Memory capacity in the Hopfield model when requiring a minimum basin size of one spin flip. a) The average and standard deviation of the number of stored memories for the $n=16$ Hopfield model versus the number of intended memory patterns $P$. Memories must exhibit basin sizes greater than one spin flip at the 50\% recall level to be counted in our definition of capacity. The results are averaged over 10,000 random realizations of the $\mathbf{J}$ matrix for each $P$. While the number of stored memories initially tracks the number intended memories, the Hebbian learning rule begins to fail at larger $P$. This leads to a maximum of approximately 3.59 memories at $P=4$. b) The average and standard deviation of the number of stored memories in the Hopfield model versus $n$ using the same analysis described in panel (a). A linear fit to the data beginning at $n=10$ (to avoid small-$n$ effects) yields an average memory capacity of $2.13+0.10\cdot n$.   }
    \label{fig:HopCap}
\end{figure}

The number of stored memories and its standard deviation versus $P$ are shown for $n=16$ in Fig.~\ref{fig:HopCap}a. We find that the maximum number of stored memories is 3.59(1), on average, and this maximum is achieved when intending to store $P=4$ memories. This is greater than the $2.21$ capacity estimated by the thermodynamic scaling relation $0.138\cdot n$, indicating that there are finite-size effects not captured by the thermodynamic bound. We repeat this procedure for up to $n=50$. The average memory capacity and its standard deviation over disorder realizations of random $\xi^p$ are shown in Fig.~\ref{fig:HopCap}b. We find the average memory capacity fits well to the form $2.13+0.10\cdot n$ for $n\gtrsim 10$, as opposed to the thermodynamic estimate $0.138\cdot n$. The offset $2.13$ is a finite-size correction that improves the estimate of the memory capacity at small $n$; the $0.138\cdot n$ estimate incorrectly predicts a memory capacity of less than one for $n<8$. The scaling $0.10\cdot n$ indicates that requiring a minimum basin size of one spin flip and permitting no deviations in the stored pattern reduces the slope of the $0.138\cdot n$ scaling estimate for Hopfield capacity.

The approximate capacity $2.13+0.10\cdot n$ is used in Figure~\ref{Fig3}b of the main text. There, the Hopfield regime is defined by memory capacities bounded by $2.13+0.10\cdot n$ on the $x$-axis, and bounded by the maximal average basin size $2^n/(2.13+0.10\cdot n)$ on the $y$-axis. These bounds define the shaded area in the plot and approximate the regime in which Hopfield networks can operate. 

\subsection{SK model spin glass memory capacity}
\label{sec:SKmodel}

A primary result of our work is the demonstration that spin glasses can operate as associative memory neural networks given a suitable form of dynamics. The SK spin glass~\cite{Sherrington1975smo}, evolving under SD dynamics, provides a simplified model of our cavity QED system that captures its ability to perform associative memory recall in a spin glass~\cite{Marsh2021eam}. In this section, we benchmark the recall capability of the SK model to provide a comparison to our experimental data.  

We have previously shown that the SK model $\mbf{J}$ matrices provide a simple approximation to the disordered $\mbf{J}$ matrices generated by multimode cavity-mediated interactions~\cite{Marsh2021eam,Marsh2025amc}. The energy of the SK model is given by $E=-\frac{1}{2}\sum_{i,j}^n J_{ij}s_i s_j$, where each $s_i=\pm1$ is a binary spin variable. Each element $J_{ij}$ of the symmetric coupling matrix is an independent and identically distributed random variable. The $J_{ij}$ are sampled from a Gaussian distribution to yield a single disorder realization of the spin glass. We consider a zero-mean distribution, corresponding to the point that is deepest in the spin glass phase~\cite{Sherrington1975smo}. 

\begin{figure}[t]
    \centering
    \includegraphics[width=0.8\linewidth]{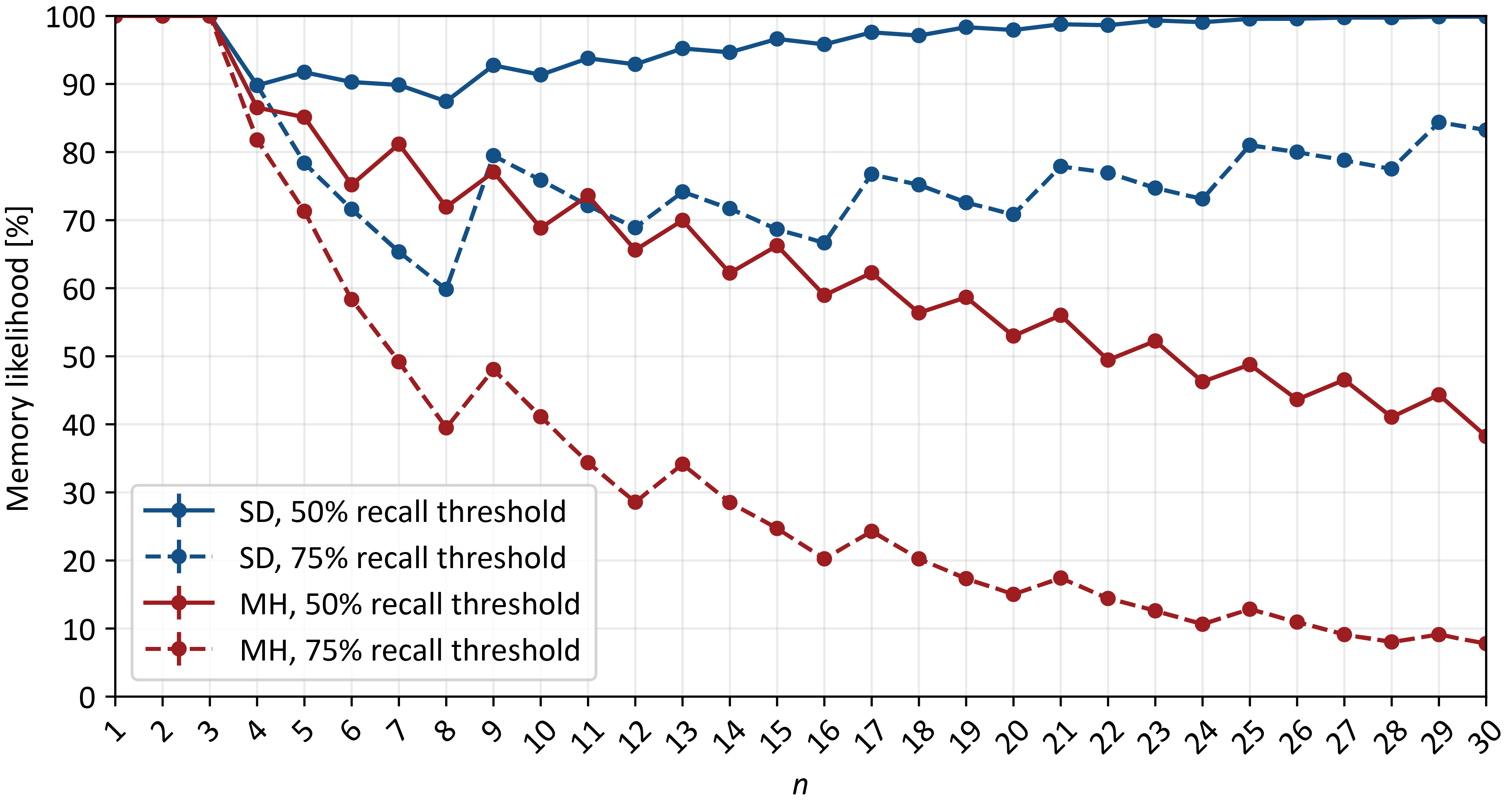}
    \caption{Fraction of local minima that are associative memories in the SK model spin glass. Each data point is averaged over at least 200 realizations of SK-model $\mbf{J}$ matrices. For each $\mbf{J}$, local minima are found by relaxing up to 54,000 random initial spin states to local minima. Standard error is smaller than the marker in all cases. Blue (red) lines show the fraction of minima that are memories using SD (MH) dynamics. Solid (dashed) lines show the results requiring a minimum basin size of one spin flip at the 50\% (75\%) recall level.}
    \label{fig:SKMemProb}
\end{figure}

Each of the exponentially many local minima for a given $\mbf{J}$ are candidate memories. As in both our experimental data and in our analysis of the Hopfield model, we require that a memory demonstrates a basin size of at least one spin flip at the 50\% recall level. We consider the SK model evolved with either equilibrium MH dynamics or SD dynamics, as described in the main text. Our previous theoretical work predicts that the SK model with MH dynamics does not function as an associative memory, while the SK model with SD dynamics does. 

We start by numerically computing the likelihood that a local minimum in the SK model exhibits a basin size greater than one spin flip. This probability is plotted versus $n$ in Fig.~\ref{fig:SKMemProb}. We find that for SD dynamics this probability approaches 100\% with increasing $n$, while it monotonically decreases with $n$ (up to a small even/odd $n$ dependence) for MH dynamics. This is consistent with the prediction of Ref.~\cite{Marsh2021eam} that for large $n$, all local minima in the SK model become memories under SD dynamics, while none do using MH dynamics. We additionally plot the fraction of memories using a more stringent 75\% recall threshold. While the trends are consistent, we note that the asymptotic approach to 100\% under SD dynamics occurs more slowly with $n$. The true cavity QED dynamics are predicted to be closer to SD than MH~\cite{Marsh2021eam}. We therefore use a 50\% recall threshold in the analysis of our experimental data to avoid these more severe finite-size effects.  

The memory capacity of the SK model under SD and MH dynamics was plotted in Fig.~\ref{Fig3}, where the capacity under MH  dynamics appears to bend upwards. However, the basin size of memories under MH dynamics is predicted to decrease to zero with increasing $n$, while a nearly extensive basin size is predicted for SD~\cite{Marsh2021eam}. Thus, the memory capacity under MH may eventually plateau or begin to decrease at larger $n$. Seeing this potential turnaround in the MH capacity falls outside the current capabilities of our numerical methods but is not required for the current study. Rather, the simulated SK memory capacity at the experimentally realized system sizes up to $n=16$ provide a direct comparison point for the experimentally measured capacities in Fig.~\ref{Fig3} of the main text.

\subsection{Simulation of quantum-optical neural networks}\label{sec:semiclassicalSims}

We now simulate associative memory recall in quantum-optical neural networks using the semiclassical model presented in Sec.~\ref{sec:semiclassical}. The semiclassical model we study yields estimates of the memory capacity and its enhancement due to polaronic effects that are consistent with our experimental data.

\begin{figure}[t]
    \centering
    \includegraphics[width=\linewidth]{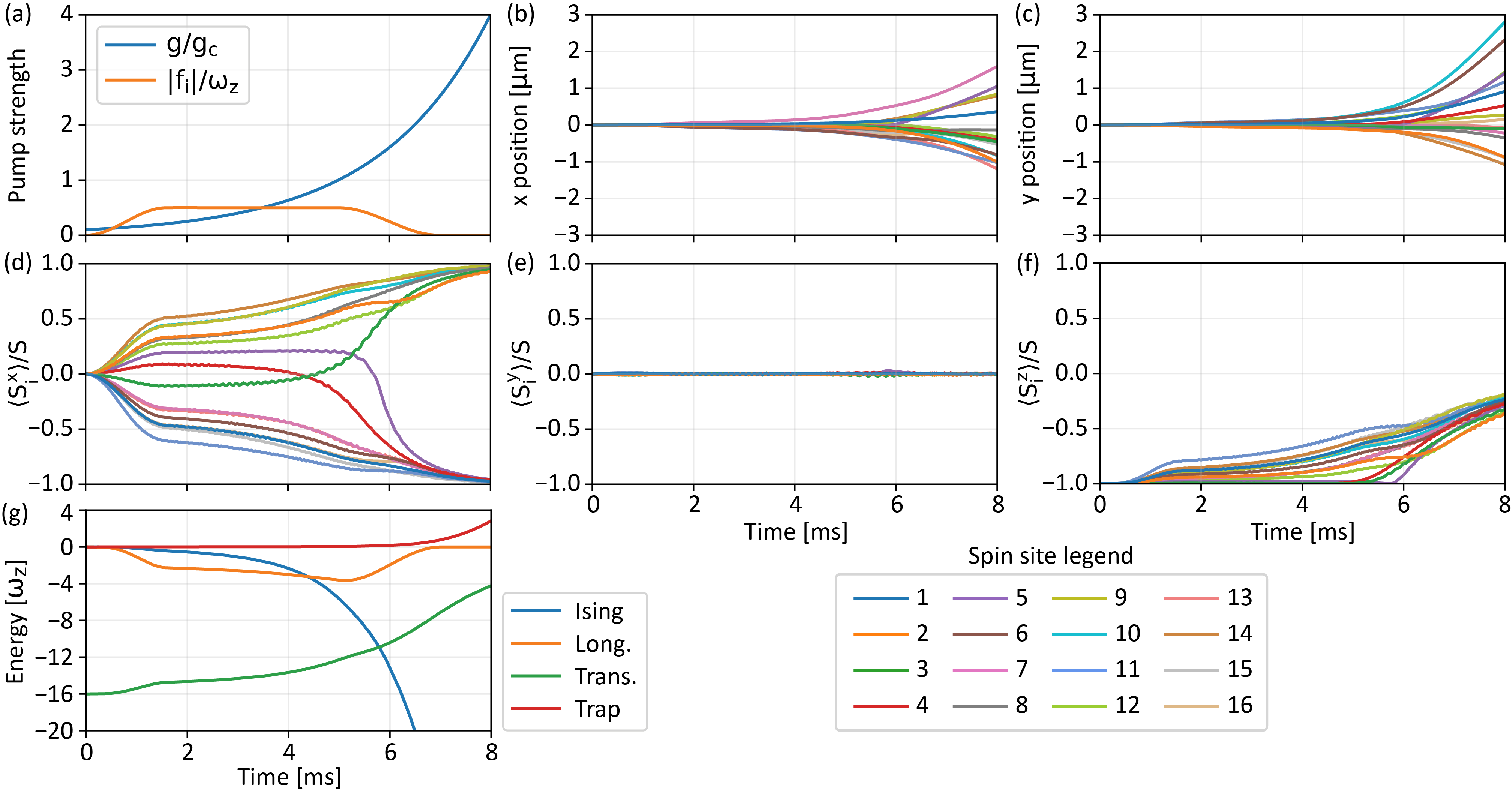}
    \caption{ Simulation of a single recall trial. a) The transverse pump strength (blue) and longitudinal pump strength (orange) follow schedules approximating the experimental sequence shown in Fig.~\ref{fig:ramps}. b) Evolution of the $x$ components of the position deviations $\mbf{r}_i-\mbf{r}_i^0=(x_i-x_i^0,y_i-y_i^0)$. c) Evolution of the $y$ components of the position deviations. d-f) Time evolution of the $\ex{\hat{S}_i^{x/y/z}}$ spin components. g) Evolution of the Ising energy $-\hbar g\sum_{ij=1}^n J(\mbf{r}_i,\mbf{r}_j)\ex{\hat{S}_i^x}\ex{\hat{S}_j^x}$, the energy due to the longitudinal fields $- \hbar\sum_{i=1}^n f(\mbf{r}_i) \ex{\hat{S}_i^x}$, the energy due to the transverse field $\hbar\omega_z \sum_{i=1}^n \ex{\hat{S}_i^z}$, and the trap energy  $E_\text{trap}\sum_{i=1}^n(\mbf{r}_i-\mbf{r}_i^0)^2/(2w_t^2)$. The energies are normalized by $S$. The Ising energy reaches approximately $-60 \omega_z$ by $t=8$~ms.}
    \label{fig:sim}
\end{figure}

A single recall trial is simulated by numerically solving the equations of motion in Eq.~\eqref{eq:EOMsemiclassical} using an 8th order Runge-Kutta method. The spins are all initialized in the normal state, $\ex{\hat{S}_i^z}=-S$ and $\ex{\hat{S_i^x}}=\ex{\hat{S_i^y}}=0$. The CoM positions $\mbf{r}_i$ are initialized at their respective trap locations $\mbf{r}_i^0$. The parameter $\omega_z$ is set to match the experimental value provided in the main text. We use a damping coefficient $c_d=\hbar/(0.002\,\mu\text{m}^2)$, corresponding to a damping ratio of approximately 1,000 for our trap frequencies. This puts the positional dynamics deep into the overdamped regime, ensuring the gradient descent approximation to the classical position dynamics is accurate. The results that follow are insensitive to the precise value of $c_d$. The trap energy $E_\text{trap}$ and trap width $w_t$ are set so that the average magnitude of position deviations due to spin-position coupling matches the experimental value of approximately 1.3~$\mu$m.  

The transverse and longitudinal pump schedules used in the simulations are shown in Fig.~\ref{fig:sim}a, which are designed to closely approximate the experimentally realized ramp schedule shown in Fig.~\ref{fig:ramps}. The transverse pump is exponentially ramped to $4g_c$, matching the experimental ramp form, but does not include the final quench to $\sim 4.5g_c$ for imaging. That brief 300~$\mu$s quench is not expected to affect the spin configuration. The longitudinal field strengths $|f_i|$ are ramped in the simulation using a simple analytic approximation to the experimentally realized schedule in Fig.~\ref{fig:ramps}b, which is complicated by experimental constraints. The simulation uses a cosine ramp from zero at time $t=0$ to $|f_i|=0.5\omega_z|\cos(\phi_i)|$ at time $t=1.6$~ms, where $\phi_i$ is the relative phase between the transverse pump and the longitudinal pump beam focused onto spin $i$. This relative phase will be discussed further below. The amplitude $0.5\omega_z$ is set to approximately match the experimentally realized amplitude. The $|f_i|$ are then held constant until $t=5$~ms when they ramp down to zero with another cosine functional form finishing at $t=7$~ms. This leaves 1~ms of spin evolution time in the absence of longitudinal pumping before measuring the spin state at $t=8$~ms, matching the experimental conditions. 

Figure~\ref{fig:sim} shows an example simulation of recall dynamics using trap locations $\mbf{r}_i^0$ that correspond to the $\mbf{J}_1$ neural network discussed in the main text. The stimulus pattern is a corrupted memory state, with randomly chosen spin flip errors applied to spins 3, 4, and 5. The spin evolution is shown in Fig.~\ref{fig:sim}d-f. The $\ex{\hat{S}_i^x}$ components are initially biased toward the stimulus fields $f_i$. However, as the transverse pump strength increases and as the stimulus fields are ramped off, spins 3, 4, and 5 flip in sign before all the spins approach a polarized $\ex{\hat{S}_i^x}$ spin configuration. These spin flips are the recall process in action; the initially corrupted spins are corrected by the neural network dynamics. Because only the initially corrupted spins flip, this is an example of a successful memory recall trial. Summarizing the evolution of the other spin components, the $\ex{\hat{S}_i^z}$ are initially polarized toward $-S$ in the normal state, but approach zero as spin organization occurs. The $\ex{\hat{S}_i^y}$ components show little evolution during recall.

The BEC positions deviate from their trap minima due to the spin-position coupling terms in Eq.~\eqref{eq:Esemiclassical}. The deviations $\mbf{r}_i-\mbf{r}_i^0$ are shown in Fig.~\ref{fig:sim}b,c. The gradient descent position dynamics minimize the total energy, including the trap energy as well as that from the spin-position coupling terms $-\hbar g \sum_{ij}J(\mbf{r}_i,\mbf{r}_j)\ex{\hat{S}_i^x}\ex{\hat{S}_j^x}$ and $-\hbar \sum_if(\mbf{r}_i)\ex{\hat{S}_i^x}$. The positions continue to evolve up to $t=8$~ms because the energy landscape continues to shift due to the increasing transverse pump strength. The positions stabilize when the transverse pump strength is held constant (not shown). Figure~\ref{fig:sim}g shows the evolution of the terms in the semiclassical energy. At $t=0$, the spins are in the normal state and the transverse term $\omega_z\sum_i\ex{\hat{S}_i^z}$ dominates the energy. As the longitudinal fields ramp up, it becomes energetically favorable for the $\ex{\hat{S}_i^x}$ to become nonzero to match the stimulus pattern. At later times, transverse pumping makes the Ising interaction dominant, and the $\ex{\hat{S}_i^x}$ organize more deeply and possibly undergo spin flips to correct initially corrupted spins in the stimulus. The trap energy begins to increase at later times, when the BECs begin to deviate significantly from their trap locations to further reduce the energy of the Ising term through spin-position coupling. 

\begin{figure}[t]
    \centering
    \includegraphics[width=0.55\linewidth]{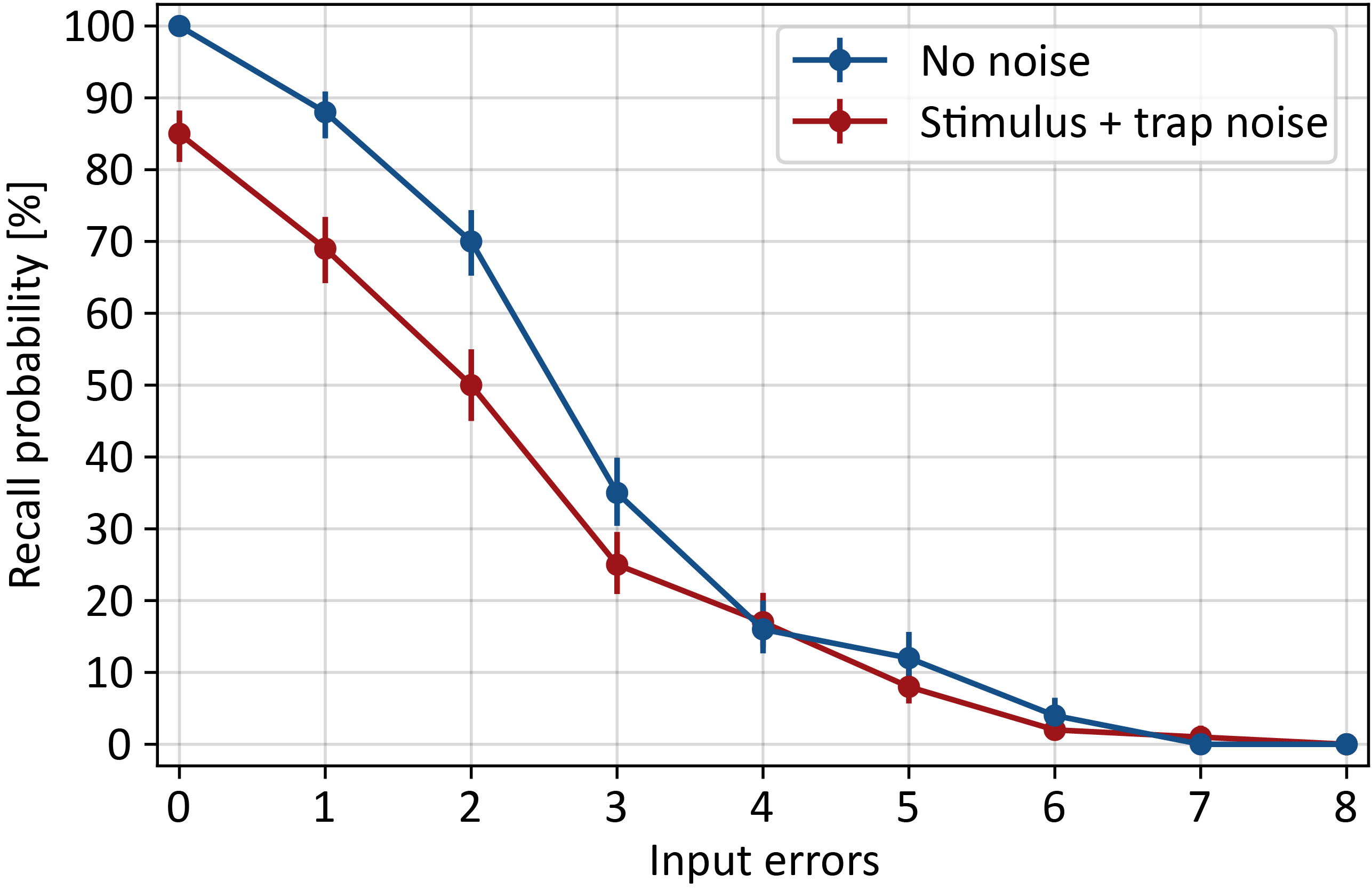}
    \caption{A typical simulated recall curve for the $\mbf{J}_1$ neural network. Each data point is averaged over 100 recall trials. The recall curve is simulated using both the experimental level of stimulus and trap noise as well as without these noise terms.}
    \label{fig:simCurve}
\end{figure}

Memories are found in the simulated system using the same technique that is used experimentally: Spin states are cataloged using ${>}400$ trials with randomly chosen stimuli. After identifying candidate memories, we test each for its recall performance. Figure~\ref{fig:simCurve} shows a typical simulated recall curve using one such memory candidate for $\mbf{J}_1$, with 100 trials per error level. We are able to include in our simulations the two dominant sources of experimental noise that can inhibit recall: trial-to-trial fluctuations in the stimulus and tweezer trap locations. As mentioned above in Sec.~\ref{sec:methods}, other sources of experimental noise, such as positional drift of the longitudinal pump beams and global atom number fluctuations, are not severe enough to have a strong effect on recall performance in our parameter regime. Fluctuations in the tweezer traps are modeled by adding Gaussian noise with standard deviation $0.5$~$\mu$m to the trap locations $\mbf{r}_i^0$ before each recall trial.  This corresponds to the experimental level of stability. Fluctuations in the stimulus consists of 10\% intensity fluctuations between different $f_i$, an uncontrolled phase $\phi_\text{trans}$ between the longitudinal and transverse pump beam, and 0.3~radian phase fluctuations between the $f_i$, as described in Sec.~\ref{sec:methods}. This means that for each recall trial, the nominal $f_i$ are replaced with $A_if_i\cos(\phi_\text{trans}+\phi_i)$ where the $A_i$ are Gaussian normal variables with mean 1 and standard deviation 0.1, $\phi_\text{trans}$ is sampled randomly from $[0,2\pi]$, and the $\phi_i$ are sampled from zero-mean Gaussian distributions with 0.3-radian standard deviation. 

Figure~\ref{fig:simCurve} shows the simulated recall curve including the experimentally measured levels of stimulus and trap noise, as well as the recall curve with these noise terms artificially turned off. The zero-error recall probability always begins at 100\% in the absence of noise, but noise can reduce it. Most, but not all of the experimental memory candidates have a zero-error recall probability below 100\% due to this effect; see Fig.~\ref{fig:caseStudy} for examples. Similarly, the noise can shrink the basin size. The severity of the reduction in recall probability varies among memories. However, the basin sizes are typically reduced more when using higher recall cutoffs for the basin size. The 50\% recall threshold we use in the main text is low enough to mitigate these effects of experimental noise. 

We numerically simulate the $\mbf{J}_1$ memory capacity in the same way that it is measured experimentally, by counting the number of memories with a basin size greater than or equal to 1 spin flip. We use the same bootstrap analysis described in Sec.~\ref{sec:AMmethods} to estimate the error. The memory capacity is simulated using three levels of trap strengths to control the level of polaronic elasticity in the positions: the default experimental trap strength, the $4\times$ weaker trap strength that experimentally yields enhanced memory capacity, and the limit of strong traps, for which the elastic response of the atoms is completely eliminated. Moreover, for each trap strength we simulate the memory capacity using the experimental levels of stimulus and trap noise, only stimulus noise, only trap noise, and with no noise. The results are summarized in Table~\ref{tab:sims}. We find that the simulated memory capacity matches the experimentally measured capacities for both the default and enhanced polaronic elasticity settings, within error. This suggests that the semiclassical theory may be an accurate predictor of the memory capacity, including polaronic enhancement, despite its inability to accurately  reproduce the specific memory states themselves; this is discussed below.

\begin{table}[t]
\centering
\begin{tabular}{|l|c|c|c|c|c| }
    \cline{2-6} 
    \multicolumn{1}{c|}{} & \thead{Experimental\\capacity} & \thead{Simulated capacity\\(stimulus + trap noise)} & \thead{Simulated capacity\\(stimulus noise)} & \thead{Simulated capacity\\(trap noise)} & \thead{Simulated capacity\\(no noise)}   \\ 
\hline 
\thead{Strong trap limit\\(no elasticity)} & --- & 10(1) & 11(2) & 14(1) &  22(1) \\ 
\hline
 \thead{Default trap strength\\(default elasticity)} & 14(1) & 12(2) & 17(2) & 23(2) & 36(2) \\ 
 \hline 
 \thead{$4\times$ weaker trap\\(enhanced elasticity)} & 25(2) & 23(3) & 27(3) & 60(4) & 60(4)\\ 
 \hline
\end{tabular}
\caption{Simulated $\mbf{J}_1$ memory capacity under different conditions. Each row corresponds to a different level of polaronic elasticity, as controlled by the trap strength. No elasticity corresponds to the limit of strong traps. The first column shows the experimentally measured capacities shown in Fig.~\ref{Fig3} in the main text. The second column shows the capacity simulated including the two dominant noise sources affecting recall, trial-to-trial fluctuations in the trap positions and stimulus field. The remaining columns show the capacities simulated using only stimulus noise, only trap noise, and without any noise. 
}
\label{tab:sims}
\end{table}

The simulations indicate that the memory capacity could be enhanced up to threefold by suppressing experimental noise sources. For example, the simulated capacity at default experimental conditions increases from 12(2) with both noise sources to 36(2) without them. In future work, fluctuations in the stimulus could be mitigated by phase locking the longitudinal and transverse pump beams. Replacing the DMD with a spatial light modulator would additionally offer a greater degree of tunability and improved power efficiency, leading to stronger and more homogeneous $f_i$. Trap position fluctuations likely result from differential movement between the cavity and optical table. This could be mitigated by, e.g., locking the tweezer positions to a fixed point in the cavity's frame of reference. 

The simulations additionally reveal elasticity to be a defense mechanism against $\mbf{J}$ chaos arising from position fluctuations. Fluctuations in the trap locations directly lead to fluctuations in the $\mbf{J}$ matrix through the position dependence described by Eq.~\eqref{eq:J}. The $\mbf{J}$ fluctuations, in turn, can negatively impact recall performance. For example, in the absence of elastic effects, the simulated capacity drops from 22(1) without noise to 14(1) when including only trap noise; the decreased capacity is due to $\mbf{J}$ chaos. However, elasticity makes the spins less sensitive to initial position misplacement by giving them the ability to move and correct their positions. The simulated capacity with enhanced elastic effects does not drop at all upon including trap noise, remaining 60(4) in both the noise-free case and when including only trap noise. Overall, the simulations reveal elasticity as a mechanism to defend against $\mbf{J}$ chaos by making $\mbf{J}$ a dynamical quantity. This could play an important role in preserving recall performance in larger-scale spin glass associative memories.

We note that the semiclassical model is not accurate enough to predict the memory states themselves with high accuracy. To quantify the agreement, we compare the simulated memories with those found experimentally. For each simulated memory, we find the experimental memory that best matches, or in other words, differs by the least number of spin flips. For $\mbf{J}_1$, the simulated memories contain an average of approximately six spin flip errors compared to its closest-matching experimental memory. The discrepancy arises most likely because of imprecision in our analytic estimate of $J(\mbf{r}_i,\mbf{r}_j)$, which does not capture experimental details such as imperfect mode degeneracy~\cite{Kroeze2023hcu}. A precise match to the experimental memories is not needed for simulating the memory capacity and polaronic effects, which appear to be captured well by the semiclassical theory.

%%%%%%%%%%%%%%%%%%%%%%%%%%%%%%%%%%%%%%%%%%%%%%%%%%%%%%%%%%%%%%%%%%

\section{Extended data}\label{sec:extendedData}

This section presents additional data supporting the conclusions in the main text. We present here a more extensive characterization of the $\mbf{J}_1$ neural network, an analysis of memory capacities using a more strict 75\% recall threshold, and a demonstration of associative memory in an $n=20$ neural network. 

\subsection{$\mbf{J}_1$ memory candidates}

The $\mbf{J}_1$ neural network pertaining to Fig.~\ref{Fig2} was characterized more fully than the others. The recall curve was measured using 30 trials per error level for each memory candidate that demonstrated at least 75\% zero-error recall probability. For other neural networks, we employ the faster basin estimation algorithm described in Sec.~\ref{sec:AMmethods}. We present these $\mbf{J}_1$ recall curves below and show that they fit well to a $\tanh$ functional form.    

\begin{figure}[t]
    \centering
    \includegraphics[width=\linewidth]{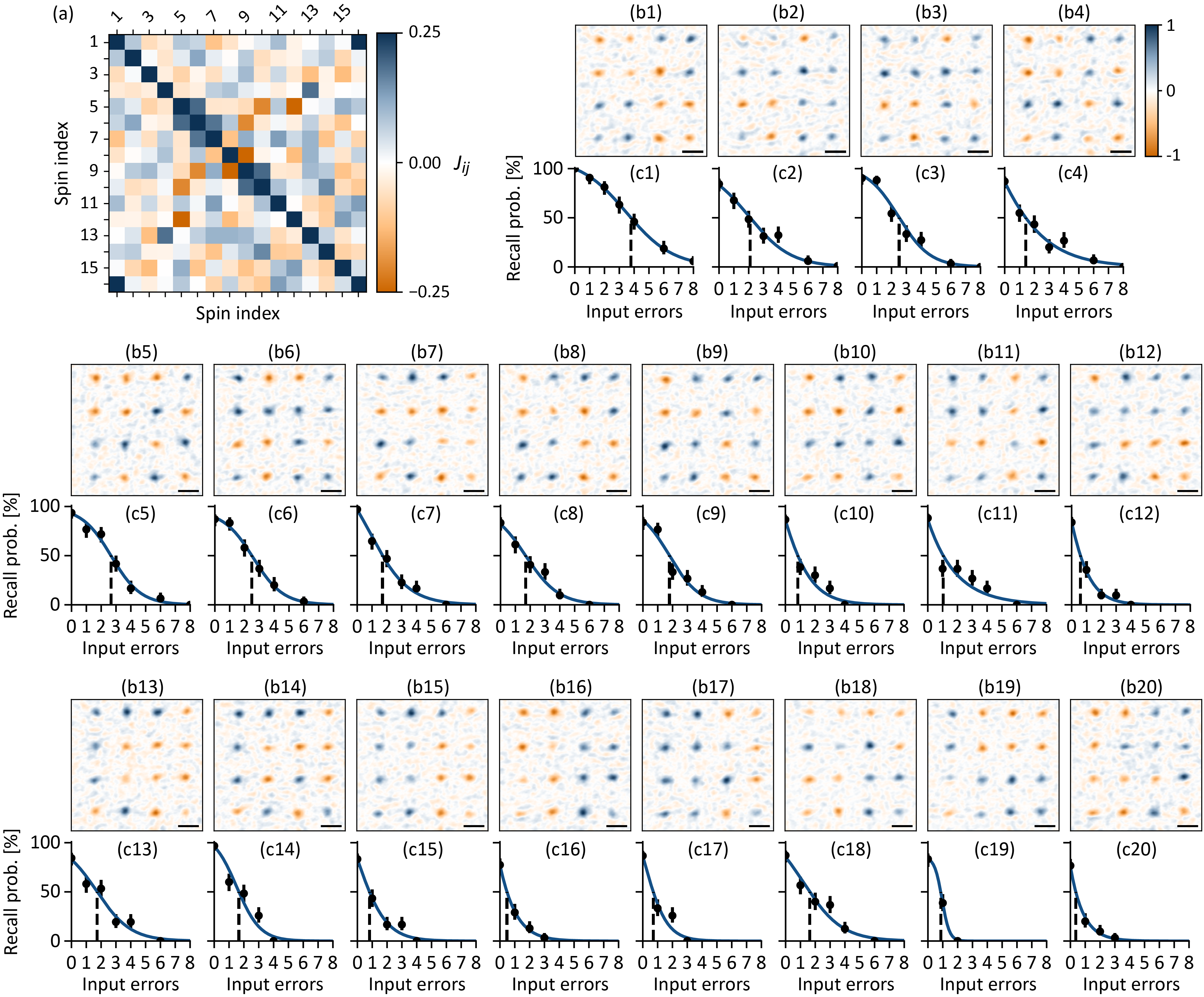}
    \caption{All $\mbf{J}_1$ memory candidates with zero-error recall probability greater than 75\%. a) Calculated $\mbf{J}_1$ matrix using Eq.~\eqref{eq:J}, demonstrating effectively random, sign-changing interactions. b1-20) Representative images for each of the twenty memory candidates. Each image is normalized to its maximum amplitude. The black bar in each image shows $w_0=34.8$~$\mu$m. c1-20) Recall curves for each memory candidate, fit to the functional form $a_1[1-\tanh(a_2x-a_3)]$.}
    \label{fig:caseStudy}
\end{figure}

Figure~\ref{fig:caseStudy} shows extended recall data for $\mbf{J}_1$. The $\mbf{J}_1$ matrix is shown explicitly in Fig.~\ref{fig:caseStudy}a and is computed via $J_{ij}=J(\mbf{r}_i^0,\mbf{r}_j^0)$ using Eq.~\eqref{eq:J}. We evaluate $J_{ij}$ using the trap locations $\mbf{r}_i^0$, corresponding to the $\mbf{J}$ matrix before spin-position coupling begins to dynamically modify $\mbf{J}$. The $\mbf{J}$ matrix is seen to fluctuate randomly in sign and strength between different spin pairs. This generates frustrated spin interactions, yielding a spin glass~\cite{Marsh2025amc}. The diagonal elements are always positive and extend beyond the range of the color bar, reaching an average value of approximately 0.7. These diagonal interactions induce a collective superradiant emission within the same atomic ensemble associated with the BEC adopting one or the other density wave pattern; in other words, these interactions encourage the effective spins within the same atomic ensemble to align. Antidiagonal matrix elements correspond to pairs of spins that are approximately at mirror image locations reflected through the cavity center. This can lead to stronger coupling strengths due to the cavity Green's function shown in Eq.~\eqref{eq:Geta}, but does not significantly affect the random nature of the coupling matrix.

Figure~\ref{fig:caseStudy} additionally shows each of the 20 memory candidates and their recall curves. The candidates are ordered according to how frequently they were encountered during the initial sampling with random stimuli. Each recall curve is fit to the functional form $a_1[1-\tanh(a_2x-a_3)]$ to estimate the basin size at the 50\% recall level. The average reduced chi-squared statistic for the $\tanh$ fit is 1.17, indicating consistency between the $\tanh$ form and the measured data. For example, memory candidate 1 in Fig.~\ref{fig:caseStudy} has a  basin size that exceeds our one-spin-flip threshold and corresponds to memory 1 in the main text. Similarly, memory candidates 5-7 in Fig.~\ref{fig:caseStudy} are labeled in the main text as memories 2-4, respectively.     

\subsection{Memory capacity using 75\% recall threshold}

We use a recall threshold of 50\% to measure the basin size and determine the memory capacities presented in the main text. This threshold is a free parameter; the minimum acceptable recall probability may depend on the specific application of the neural network. However, we show in Sec.~\ref{sec:SKmodel} that higher recall thresholds yield more severe finite-size effects that reduce memory capacity. We also show in Sec.~\ref{sec:semiclassicalSims} that using higher recall thresholds makes the memory capacity more susceptible to experimental noise sources. We thus use a 50\% threshold because it mitigates finite-size effects and reduces the impact of experimental noise while still being high enough to be considered a reasonable threshold for acceptable recall performance. Nevertheless, we present the memory capacities using a more strict 75\% threshold for all measured neural networks in this section.

\begin{figure}[t]
    \centering
    \includegraphics[width=0.8\linewidth]{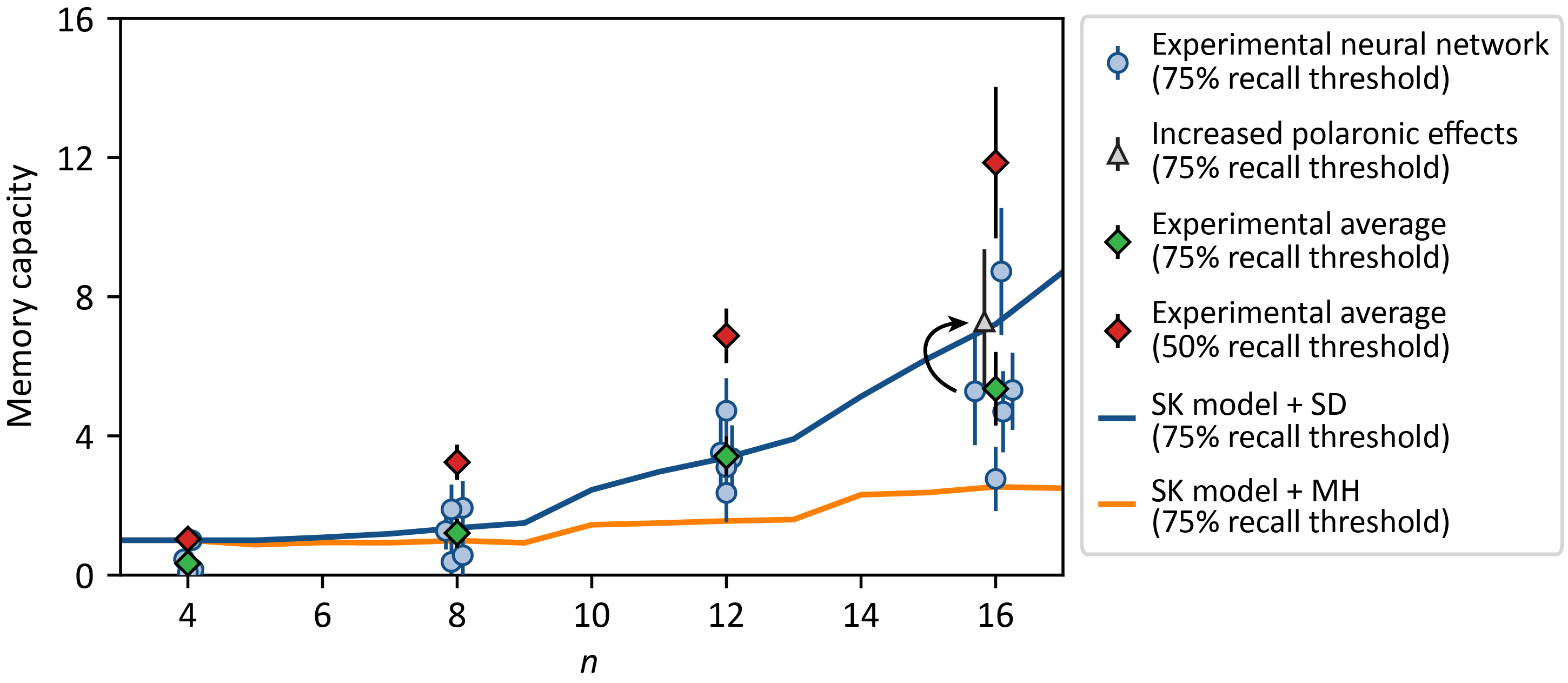}
    \caption{Experimental memory capacities using a 75\% recall threshold. The capacity for each experimentally realized neural network is shown with a blue circle and error bar. The $n=16$ neural network with enhanced polaronic elasticity is shown with a gray triangle. The arrow points from the $\mbf{J}_1$ neural network to its more elastic version. The experimental average using the 75\% recall threshold is shown using green diamonds. The average capacity with the 50\% recall threshold used in the main text is reproduced using red diamonds for comparison. The average simulated capacity of the SK model using SD (MH) dynamics is shown with a blue (orange) line.}
    \label{fig:capacity75}
\end{figure}

Memory capacities using a 75\% recall threshold are determined in the same manner as for the 50\% threshold. The basin size is estimated for each candidate memory, but the basin size now corresponds to the intersection of the recall curve with the 75\% recall threshold, leading to smaller basin sizes. The memory capacity is still the number of memories with a basin size greater than or equal to one spin flip. Error bars are determined via the same bootstrap analysis described in Sec.~\ref{sec:AMmethods}. The results are shown in Fig.~\ref{fig:capacity75}. As expected, the capacity drops across all $n$. The average capacity at $n=16$ drops from 11.9(6) to 5(1). The neural network with enhanced polaronic effects still yields an increase in capacity, but the increase is now smaller. The decrease in memory capacity using the 75\% threshold is most severe for the more elastic neural network. This is because it has the most memory candidates with a basin size near one spin flip, and so it loses the most memories when enforcing the more stringent recall requirement.    

The memory capacity in the SK model is also reduced when using the 75\% threshold. This is computed numerically in the same way as described in Sec.~\ref{sec:SKmodel} but using a 75\% recall threshold. The average memory capacity is plotted for both SD and MH dynamics in Fig.~\ref{fig:capacity75} for comparison with the experimental neural networks. The average experimental capacity remains roughly consistent with the SK model under SD dynamics; the lower capacity at $n=16$ may be due to statistical fluctuations or experimental noise not present in the SK model simulations. Overall, while the memory capacity decreases using a 75\% recall threshold, this decrease is as expected and is due to finite-size effects and increased susceptibility to experimental noise.  The distinction remains with respect to the Hopfield capacity.

\subsection{A 20-site neural network}

We present an initial demonstration of associative memory in an $n=20$ neural network. The experimental system is able to realize neural networks with $n=20$ spin ensembles and perform recall sequences as described in Sec.~\ref{sec:methods}. However, power limitations on the longitudinal pump lead to a 10\% reduction in the stimulus strength $|f_i|$. This is a minor effect, and so the same experimental recall procedure is retained for $n=20$. 

\begin{figure}[t]
    \centering
    \includegraphics[width=1.0\linewidth]{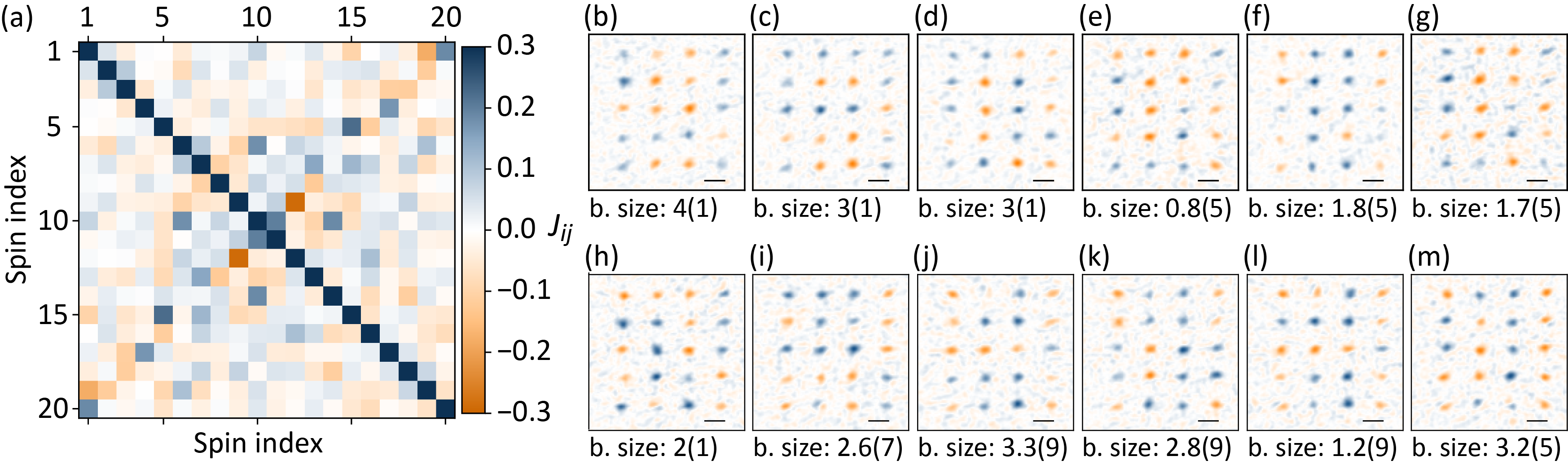}
    \caption{ Memories in an $n=20$ neural network. a) The $\mbf{J}$ matrix as computed using Eq.~\eqref{eq:J}. The diagonal elements extend beyond the range of the color bar and have an average value of approximately 0.5. Antidiagonal matrix elements correspond to spins at approximately mirror image locations in the cavity and have slightly stronger coupling strengths, as seen in Fig.~\ref{fig:caseStudy}a for the $\mbf{J}_1$ neural network. b-m) The first 12 memory candidates found with basin sizes close to or greater than one spin flip. Basin sizes estimated using the algorithm described in Sec.~\ref{sec:AMmethods}.
    }
    \label{fig:n20}
\end{figure}

Full characterization of the memory capacity becomes prohibitively time-consuming at $n=20$ due to the rapid increase in the number of memory candidates. We encountered 97 memory candidates after performing approximately 1,100 trials with random stimuli. The recall performance was measured for the 25 candidates that were found most frequently in the random sampling. Of the 25 candidates we tested, 11(1) are assessed to possess basin sizes greater than or equal to one spin flip and therefore satisfy our criterion as memories. The top 12 memory candidates are shown in Fig.~\ref{fig:n20} along with their estimated basin size using the algorithm described in Sec.~\ref{sec:AMmethods}. Thus, the memory capacity of this $n=20$ neural network is bounded from below by 11(1) but is likely much higher, since only 25 of the 97 memory candidates were tested. Properly characterizing the abundance of memories in neural networks with $n>16$ requires methods beyond the brute-force searching and testing methods employed in this work. Future work could explore stochastic sampling of memories, or implement multiple recall trials per experimental sequence, to better investigate associative memory in larger-scale spin glasses.

\end{document}